\documentclass[final,3p,times,onecolumn]{elsarticle}
\usepackage{graphicx}
\usepackage{amssymb}
\usepackage{amsmath}
\usepackage{hyperref}
\usepackage{color}
\usepackage{gensymb}

\journal{Commun. Nonlinear Sci. Numer. Simul.}
\begin{document}

\title{Effect of quartic-quintic beyond-mean-field interactions on a  self-bound dipolar droplet }

\author[fac]{Luis E. Young-S.}
\ead{lyoung@unicartagena.edu.co}
\cortext[author]{Corresponding author.}
 \address[fac]{Grupo de Modelado Computacional, Facultad de Ciencias Exactas y Naturales, Universidad de Cartagena,\\ 130014 Cartagena, Bolivar, Colombia}
\author[int]{S. K. Adhikari\corref{author}}
\ead{sk.adhikari@unesp.br}
\ead[url]{professores.ift.unesp.br/sk.adhikari/}
 \address[int]{Instituto de F\'{\i}sica Te\'orica, UNESP - Universidade Estadual Paulista, 01.140-070 S\~ao Paulo, S\~ao Paulo, Brazil}

\begin{abstract}

 We study the effect of beyond-mean-field  quantum-fluctuation (QF) Lee-Huang-Yang (LHY)  and three-body  interactions, with quartic and quintic nonlinearities, respectively,   on the formation of a stable self-repulsive (positive scattering length $a$) and a self-attractive (negative $a$) self-bound dipolar Bose-Einstein condensate (BEC)
droplet in free space  under the action of two-body contact  and dipolar 
interactions.  Previous studies of dipolar droplets
considered either the LHY interaction or the three-body interaction, as  
either of these interactions  could stabilize a  dipolar  BEC droplet against collapse. We find that the effect of three-body recombination on the formation of a dipolar droplet could be quite large  and for a complete description of the problem both the QF LHY {\it and} three-body interactions  should be considered simultaneously, where appropriate. In the self-repulsive case for small $a$ and in the 
self-attractive case, no appropriate LHY interaction is known and only three-body interaction should be used, otherwise both beyond-mean-field  interactions should be used. 
We consider a  numerical solution of a highly-nonlinear beyond-mean-field model  as well as a  variational
approximation to it in this investigation and present results for  size, shape and energy of a dipolar  droplet of polarized $^{164}$Dy atoms. The shape is  filament-like, along the polarization direction, and could be long, for a large number of atoms $N$, short for small $N$, thin for negative $a$ and small positive $a$, and fat for large positive $a$.

\end{abstract}


\maketitle

\section{Introduction}

 An one-dimensional (1D)   bright soliton remains  bound due to a balance between
defocusing forces and nonlinear attraction and  can travel at a constant velocity \cite{cite1} without deformation.
Solitons have been found 
on water surface,  in nonlinear optics  \cite{cite2} and in Bose-Einstein condensate
(BEC) \cite{cite3}. An 1D integrable analytic soliton with strict momentum
and energy conservation guarantees 
shape preservation. In a self-attractive ($a< 0$) cigar-shaped BEC with atomic scattering length $a$, quasi-1D solitons 
have been observed \cite{cite3} by applying  strong confining traps in transverse directions, following
a theoretical suggestion \cite{cite4}. A BEC soliton with only attractive  contact interaction cannot be realized \cite{towne} in  two (2D)
and three (3D) dimensions due to a collapse instability. 

Nevertheless, there has been intense research activity on a self-bound BEC in free space. 
In 2D and 3D, a self-bound spinor BEC can be formed in the presence of a spin-orbit coupling interaction \cite{SO-sol}.   
The spin-orbit
coupling can  generate an effective interatomic repulsion at short distances 
cancelling the mean-field attraction and stabilize the  spinor BEC
against collapse \cite{saka}.
In 3D,  the  inclusion of a  quantum-fluctuation (QF) Lee-Huang-Yang (LHY)  
interaction \cite{lhy} or of  repulsive three-body interaction \cite{3bdth}, in the beyond-mean-field   model of a BEC, 
can stop the collapse and thus  produce  a self-bound BEC \cite{3bd}, 
self-bound binary BEC \cite{petrov} and a self-bound binary Bose-Fermi superfluid mixture \cite{ad-bf}. 
The LHY interaction in the binary mixture
arises due to the QF correction appropriate to the repulsive intraspecies interaction \cite{petrov}. 
Self-bound binary BECs of $^{39}$K  atoms  have  been observed    \cite{obser}.  The role of the three-body interaction in atomic systems have been emphasized from an experimental \cite{3bdex} point of view as well as in self-bound condensates \cite{3bd}, harmonically-trapped condensates \cite{3bdtomio} and in condensates on optical lattices \cite{3bdol}.


In a different front,
high-density droplets  
were observed in a  trapped strongly  dipolar BEC of $^{164}$Dy \cite{19,1d1}
and $^{168}$Er \cite{20} atoms. 
In the framework of a beyond-mean-field model, the QF LHY interaction, appropriately modified for dipolar atoms, 
 \cite{qf1} can stabilize \cite{santos}
such a strongly dipolar BEC droplet against collapse. 
A self-bound  dipolar BEC can also be formed in free space under the action of the QF  LHY interaction in the self-repulsive case ($a>0$) \cite{blakie1,blakie2,drop3} or under the action  of the three-body interaction   \cite{adhidrop}.  Higher-order nonlinearities can also stabilize a self-attractive  nondipolar BEC and form a self-bound droplet in 3D \cite{sabri}.
A 3D self-bound BEC, although bears some similarity with a soliton, does not satisfy the strict energy and momentum conservation laws of an analytic 1D soliton and is called a droplet.

In this paper we study the formation of a self-bound self-repulsive  and  self-attractive   dipolar droplet of  $^{164}$Dy atoms, polarized along the $z$ direction,  including both the QF LHY and three-body  interactions. Presently, the QF LHY  interaction for dipolar atoms, appropriate for the calculation of stationary states,  is ``known'' \cite{qf1}
only in the self-repulsive case for dipolar length $a_{\mathrm{dd}}$, viz. (\ref{add}), satisfying   $a_{\mathrm{dd}} < a$, where this interaction is real,  or for 
  $a_{\mathrm{dd}}   \gtrapprox a$, where this interaction is complex with a negligible imaginary part. 
In the domain $a_{\mathrm{dd}} < a$ the system is too repulsive and no self-bound droplet can be formed.
For 
$a_{\mathrm{dd}}   \gg a$, the imaginary part becomes  large  and the present form of LHY interaction \cite{qf1} should not be used for the calculation of stationary droplet states \cite{review}. 
Nevertheless, for $a_{\mathrm{dd}}   \gg a$,  a real three-body interaction alone can lead to a self-bound dipolar droplet and will be employed in this study.
We find that the three-body interaction may have a substantial effect on the formation of a droplet, specially, on its energy, shape  and size,  for small  $a$. The dipolar droplets have a  filament-like shape, along $z$ direction, which can be long for large $N$ and short for small $N$, fat for large positive $a$, and thin for negative $a$ and small positive $a$.  
In previous studies \cite{santos,blakie1,blakie2} only long filament-like droplets  for large $N$  were found and highlighted in the self-repulsive case
stabilized by the QF LHY interaction alone.
The QF LHY and three-body  interactions are the two beyond-mean-field interaction of lowest and next-to-lowest 
orders, with quartic and quintic nonlinearities, respectively,  and for a complete description of the formation of  a self-bound  dipolar droplet, we will consider    both these interactions, where appropriate, in both self-repulsive and self-attractive cases.   For large $a$ ($a_{\mathrm{dd}}/4 \lessapprox a<a_{\mathrm{dd}}$) we consider both QF LHY and three-body  interactions and for small $a$ ($a  \lessapprox a_{\mathrm{dd}}/4 $) only the three-body interaction is considered.
 We consider a numerical solution of the underlying beyond-mean-field model including the QF LHY interaction and a  repulsive three-body interaction. For an analytic understanding of the results we also employ a variational approximation to the beyond-mean-field model.


In Sec. \ref{II} we present the beyond-mean-field model including the QF LHY   and  three-body interactions as well as  an analytic  variational approximation to this model using a Gaussian ansatz for the wave function. The use of this ansatz leads to an analytic approximation to energy
and a minimization of this energy with respect to the widths of the wave function fixes the energy as well as  the widths of the self-bound droplet.
In Sec. \ref{III} we present the variational  results for energy and  root-mean-square (rms) size  of  3D self-bound BEC droplets of $^{164}$Dy atoms and compare these with  the same obtained from a numerical solution of the beyond-mean-field model.
Finally, in Sec. \ref{IV} we present a summary of our findings.

\section{Beyond-Mean-field model}

\label{II}

In this paper we base our study on a  3D beyond-mean-field model,  including the quantum-flucuation LHY interaction  and three-body repulsion, for a self-bound dipolar droplet. 
We consider a  BEC of $N$ dipolar $^{164}$Dy atoms  polarized along the $z$ axis, of mass $m$ each,
interacting through the following 
atomic dipolar and contact interactions   \cite{dipbec,dip,yuka}
\begin{eqnarray}
V({\bf R})= 
\frac{\mu_0 \mu^2}{4\pi}\frac{1-3\cos^2 \theta}{|{\bf R}|^3}
+\frac{4\pi \hbar^2 a}{m}\delta({\bf R }),
\label{eq.con_dipInter} 
\end{eqnarray}
where $\mu_0$ is the permeability of vacuum,  $\mu$ is the magnetic dipole moment of an atom,
${\bf R}= {\bf r} -{\bf r}'$ is the vector joining two dipoles placed at ${\bf r} \equiv \{x,y,z\}$ and ${\bf r'} \equiv \{x',y',z'\}$
and $\theta$ is the angle made by  ${\bf R}$ with the  
$z$ axis. To compare the dipolar interaction with the contact interaction, it is convenient to express 
the strength of dipolar 
interaction $ \mu_0 \mu^2/4\pi$ in terms of   the  dipolar length: 
\begin{align}\label{add}
a_{\mathrm{dd}}\equiv \frac{\mu_0 \mu^2 m }{ 12\pi \hbar ^2}.
 \end{align}
The dimensionless ratio 
 \begin{equation}
\varepsilon_{\mathrm{dd}}\equiv \frac{a_{\mathrm{dd}}}{a} 
\end{equation} 
then determines
the relative strength of dipolar interaction compared to contact interaction 
and controls many properties of a dipolar BEC.
 For the formation of a self-bound dipolar droplet, the dipolar length $a_{\mathrm{dd}}$  should necessarily be greater than the scattering length: $  a_{\mathrm{dd}}>a$, thus requiring a strongly dipolar atom, where the dipolar interaction dominates over the contact interaction to make the system attractive. 
 Hence in this study we employ  $^{164}$Dy atoms with large dipole moment 
$\mu =10 \mu_{\mathrm B}$, where $\mu_{\mathrm B}$ is the Bohr magneton. The dipolar length of $^{164}$Dy atoms
is $a_{\mathrm{dd}}=130.8a_0$ and is larger than its experimental scattering length  $a=(92\pm 8)a_0$ \cite{expdy}, where $a_0$ is the Bohr radius. Consequently, a BEC of $^{164}$Dy atoms can naturally host a droplet in free space without any tuning of the scattering length to an appropriate value using a Feshbach resonance as usually required in engineering a soliton in nondipolar atoms \cite{cite3,obser}. However,  in this study we will consider the whole domain $a_{\mathrm{dd}}>a$ where a droplet can be formed.

A {\it trapless}  dipolar BEC droplet is described by the following  3D beyond-mean-field Gross-Pitaevskii (GP) equation \cite{dipbec,dip,muntsa} including the QF LHY interaction\cite{yuka,blakie} and a repulsive  three-body interaction 
\begin{align}\label{eq.GP3d}
 i \hbar  \frac{\partial  \psi({\bf r},t)}{\partial t} =&\
{\Big [}  -\frac{\hbar^2}{2m}\nabla^2
+ \frac{4\pi \hbar^2}{m}{a} N \vert \psi({\bf r},t) \vert^2 
+\frac{3\hbar^2}{m}a_{\mathrm{dd}}  N
\int \frac{1-3\cos^2 \theta}{|{\bf R}|^3}
\vert\psi({\mathbf r'},t)\vert^2 d{\mathbf r}'  
\nonumber 
 \\
& +\frac{\gamma_{\mathrm{QF}}\hbar^2}{m}N^{\frac{3}{2}}
|\psi({\mathbf r},t)|^3+\frac{\hbar N^2K_3}{2}|\psi({\mathbf r},t)|^4
\Big] \psi({\bf r},t), 
\end{align}
where  $K_3$ is the strength of the  three-body interaction  with quintic nonlinearity  and $\gamma_{QF}$ is the strength of the QF LHY interaction with quartic nonlinearity, and 
the wave function is  normalized as $\int \vert \psi({\bf r},t) \vert^2 d{\bf r}=1$. The three-body coefficient $K_3$ is small and actually complex. The imaginary part of this term, not considered here, and usually considered in other studies \cite{3bdyexp}, is responsible for a loss of atoms from the condensate due to molecule formation by  three-body recombination. The importance of the real part of this term has been emphasized in different investigations  and used in the study of harmonically-trapped condensates \cite{3bdtomio}, self-bound 3D droplets \cite{3bd}
 and condensates on optical lattices \cite{3bdol}. In this study we will ignore the imaginary part of  $K_3$ and take $K_3$ to be real and {\it positive} (repulsive)
as we will be concerned here with a stationary state and not a decaying nonstationary state, where an imaginary $K_3$ is appropriate.
The real  $K_3$, considered here, with a higher-order quintic nonlinearity, can stabilize  \cite{3bd} a self-bound  dipolar droplet against collapse in free space.

\begin{figure}[t!]
\begin{center}
\includegraphics[width=.6\linewidth]{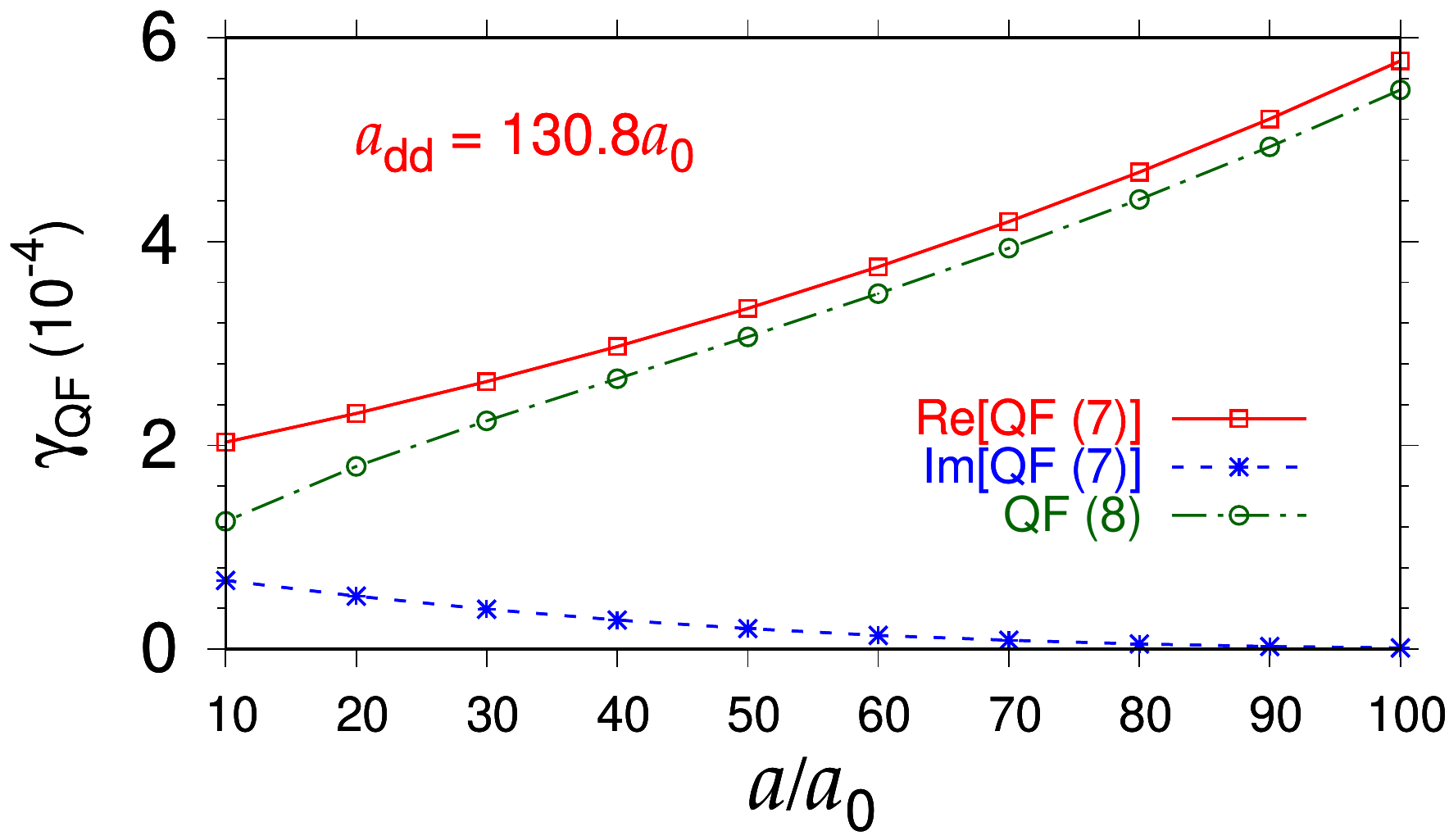}

\caption{Real (Re) and imaginary (Im)  parts of the exact LHY QF coefficient   (\ref{eq7}) and its real approximation (\ref{eq8}) versus scattering length $a$ for dipolar $^{164}$Dy atoms. Plotted quantities   are
dimensionless and the physical unit for $^{164}$Dy atoms can be restored using the unit
of length $l = 1$ $\mu$m. }

\label{fig1} 
\end{center}
\end{figure}

 The coefficient 
of the beyond-mean-field LHY-type quantum correction term $\gamma_{\mathrm{QF}}$ is given by \cite{qf1,blakie}
\begin{align}\label{qf}
\gamma_{\mathrm{QF}}= \frac{128}{3}\sqrt{\pi a^5} Q_5(\varepsilon_{\mathrm{dd}}),
\end{align}
where the auxiliary function
\begin{equation}
 Q_5(\varepsilon_{\mathrm{dd}})=\ \int_0^1 dx(1-x+3x\varepsilon_{\mathrm{dd}})^{\frac{5}{2}} .
\end{equation}
 can be evaluated to yield an analytic expression for the QF coefficient \cite{blakie}
\begin{align}\label{eq7}
\gamma_{\mathrm{QF}}=&\ \frac{128}{3}\sqrt{\pi a^5} 
\frac{(3\varepsilon_{\mathrm{dd}})^{\frac{5}{2}}}{48}  \left[(8+26\epsilon+33\epsilon^2)\sqrt{1+\epsilon}\right.
\left.
\ 15\epsilon^3 \mathrm{ln} \left( \frac{1+\sqrt{1+\epsilon}}{\sqrt{\epsilon}}\right)  \right], \quad  \epsilon = \frac{1-\varepsilon_{\mathrm{dd}}}{3\varepsilon_{\mathrm{dd}}}, \\
 \approx & \frac{128}{3}\sqrt{\pi a^5} \left( 1+\frac{3}{2} \varepsilon_{\mathrm{dd}}^2\right) .  \label{eq8}
\end{align}
  Actually, the function   $Q_5$ as well as the coefficient $\gamma_{\mathrm{QF}}$, representing a correction for dipolar atoms,  is complex for $\varepsilon_{\mathrm{dd}}>1$ and, for studies of stationary states, expression (\ref{eq7}) is formally meaningful  for $\varepsilon_{\mathrm{dd}}\le 1$, where this expression is real   \cite{review}. However its imaginary part remains small compared to its real part for medium values of $a$ where 
 $4\gtrsim \varepsilon_{\mathrm{dd}}> 1$ and  will be neglected in this study  of stationary self-bound states as in other studies, consistent with the neglect of the imaginary part of the three-body coefficient $K_3$. 
For nondipolar atoms $\varepsilon_{\mathrm{dd}}=0$, while $Q_5(\varepsilon_{\mathrm{dd}})  =1$, and one recovers the well-known LHY limit $\gamma_{\mathrm{QF}}=128\sqrt{\pi a^5}/3$ \cite{lhy} valid for repulsive hard-sphere nondipolar atoms. 
The approximation (\ref{eq8}) to QF coefficient, advocated in Ref. \cite{blakie}  for small $\varepsilon_{\mathrm{dd}}$, has often been used in different studies \cite{blakie1,blakie2,pohl,pfau} for $\varepsilon_{\mathrm{dd}} >1$.   In Fig. \ref{fig1} we display the real and imaginary parts of the 
 QF coefficient (\ref{eq7}) and its real approximation (\ref{eq8}) for different values of $a$. 
From Fig. \ref{fig1}, we find that  the agreement between the two forms QF coefficients 
is satisfactory for larger values of $a$  ($a\gtrapprox 30a_0$), where the QF LHY interaction has a small imaginary part. In this domain,
the QF LHY interaction 
will be used in conjunction with the three-body interaction in the study of stationary dipolar self-bound droplets. 
For smaller values of $a$ 
($a\lessapprox 30a_0$), where the imaginary part of the LHY interaction is large and  in the self-attractive case, where the QF interaction is not known,
we will use only the three-body interaction in the study of the self-bound dipolar  droplets.



Equation (\ref{eq.GP3d}) can be reduced to 
the following  dimensionless form by scaling lengths in units of a fixed length scale $l = 1$ $\mu$m, time in units of $ml^2/\hbar$,  density $|\psi|^2$ in units of $l^{-3}$, $K_3$ in units of $\hbar l^4 /m$, and energy in units of $\hbar^2/ml^2$
\begin{align}
i \frac{\partial \psi({\bf r},t)}{\partial t} &=
{\Big [}  -\frac{1}{2}\nabla^2
+3a_{\mathrm{dd}}  N
\int \frac{1-3\cos^2 \theta}{|{\bf R}|^3}
\vert\psi({\mathbf r'},t)\vert^2 d{\mathbf r}' 
+ 4\pi{a} N \vert \psi({\bf r},t) \vert^2
+\gamma_{\mathrm{QF}}N^{\frac{3}{2}}
|\psi({\mathbf r},t)|^3 \nonumber \\&
+
\frac{K_3N^2}{2}|\psi({\mathbf r},t)|^4
\Big] \psi({\bf r},t).
\label{GP3d}
\end{align}
Equation (\ref{GP3d})  
can also be obtained by applying the variational rule
\begin{align}
i \frac{\partial \psi}{\partial t} &= \frac{\delta E}{\delta \psi^*} 
\end{align}
with the following energy functional (energy per atom) of a stationary dipolar droplet
\begin{align}\label{en}
E &= \int d{\bf r} \Big[ \frac{|\nabla\psi({\bf r})|^2}{2}+ 2\pi Na |\psi({\bf r})|^4 
 + \frac{3}{2}a_{\mathrm{dd}}N|\psi({\bf r})|^2 
\left.  \int \frac{1-3\cos^2\theta}{R^3}|\psi({\bf r'})|^2 d {\bf r'} \right. 
\nonumber \\&
 +\frac{2\gamma_{\mathrm{QF}}}{5} N^{\frac{3}{2}}
|\psi({\bf r})|^5    +\frac{K_3N^2}{6}|\psi({\bf r})|^6
\Big].
\end{align}
For a self-bound droplet 
this energy has to be negative necessarily.

The formation of a self-bound  dipolar droplet can be understood by an analytic  variational
approximation  obtained with the following normalized axisymmetric Gaussian ansatz for the stationary wave function: 
\begin{align}
\psi({\bf r})=\frac{\pi^{-\frac{3}{4}}}{\sqrt{w_z}w_
\rho}\exp\left[-\frac{\rho^2}{2w_
\rho^2}-\frac{z^2}{2w_z^2}\right],
\end{align}
where $\rho \equiv \{x,y \}$, $w_z$ and $w_\rho$ are the widths of the Gaussian wave function. With this function, 
the energy integral  (\ref{en}) can be evaluated to yield the following analytical result \cite{dip}
\begin{align}\label{energy}
E  =&\frac{1}{2w_
\rho^2}+\frac{1}{4w_z^2}+\frac{N[a-a_{\mathrm{dd}}f(\kappa)]}{\sqrt{2\pi}w_
\rho^2w_z} +{\left(\frac{2}{5}\right)}^{\frac{5}{2}}\frac{ \gamma_{\mathrm{QF}} N^{\frac{3}{2}}}{\pi^{\frac{9}{4}}w_
\rho^3w_z^{\frac{3}{2}}}
+ \frac{K_3N^2\pi^{-3}}{18\sqrt 3 w_\rho^4 w_z^2}, \quad \kappa =\frac{ w_\rho}{w_z},
\end{align}
where
\begin{align}
f(\kappa)&= \frac{1+2\kappa^2-3\kappa^2d(\kappa)}{1-\kappa^2}, \\
d(\kappa)&= \frac{\mathrm{atanh} \sqrt{1-\kappa^2}}
{\sqrt{1-\kappa^2}}.
\end{align}
In Eq. (\ref{energy}), the first two terms on the right hand side are contributions of the kinetic energy of an atom in the droplet, the third term on the right hand side corresponds to the net attractive atomic interactions
 responsible for the formation of a self-bound  droplet and the last two terms are contributions of the  beyond-mean-field QF LHY and repulsive three-body interactions, respectively.  
 The higher order quartic and quintic nonlinearities of the  QF LHY
and three-body interactions compared to the cubic nonlinearity of the two-body interaction, has led to a
more singular repulsive term at the center  ($w_\rho$, $w_z \to 0$)  in (\ref{energy}). This makes the system highly repulsive
at the center and stops the collapse
stabilizing the self-bound dipolar  droplet.

A minimization of  energy (\ref{energy}) with respect to widths $w_\rho$  ($\partial E/\partial w_\rho=0$) and $w_z$ 
($\partial E/\partial w_z=0$) determines  the widths of the self-bound dipolar BEC
\begin{align} \label{w1}
\frac{1}{w_\rho^3}&+ \frac{N[2a-a_{\mathrm{dd}}g(\kappa)]}{\sqrt{2\pi}w_\rho
^3w_z} 
+{\left(\frac{2}{5}\right)}^{\frac{5}{2}} \frac{ 3\gamma_{\mathrm{QF}} N^{\frac{3}{2}}}{\pi^{\frac{9}{4}}w_
\rho^4w_z^{\frac{3}{2}}}   + \frac{4K_3N^2}{18\sqrt 3 \pi^3  w_\rho^5w_z^2}=0,
\\
  \label{w2}
\frac{1}{w_z^3}&+ \frac{2N[a-a_{\mathrm{dd}}c(\kappa)]}{\sqrt{2\pi}w_\rho
^2w_z^2}
+{\left(\frac{2}{5}\right)}^{\frac{5}{2}} \frac{ 3\gamma_{\mathrm{QF}} N^{\frac{3}{2}}}{\pi^{\frac{9}{4}}w_
\rho^3w_z^{\frac{5}{2}}}+ \frac{4K_3N^2}{18\sqrt 3 \pi^3  w_\rho^4w_z^3}=0,
\end{align}
where
\begin{align}
g(\kappa)&= \frac{2-7\kappa^2-4\kappa^4+9\kappa^4d(\kappa)}{(1-\kappa^2)^2},\\
c(\kappa)&= \frac{1+10\kappa^2-2\kappa^4-9\kappa^2d(\kappa)}{(1-\kappa^2)^2}.
\end{align}
The widths, obtained from a solution of Eqs. (\ref{w1}) and (\ref{w2}),  when substituted in Eq. (\ref{energy}), determine the corresponding energy.

\section{Numerical Results}

\label{III}

We  solve 3D  partial differential
 equation (\ref{GP3d}) for a dipolar BEC  numerically 
by the split-time-step Crank-Nicolson
method \cite{crank} employing the  imaginary-time propagation rule.
 There are  FORTRAN/C  programs \cite{dip}
  and their open-multiprocessing  versions  \cite{omp} approprite for  this purpose.
Often, the self-bound dipolar droplet is highly elongated along the $z$ direction; in such cases it is appropriate to take a larger number of discretization points  along the $z$ direction as compared to $x$ and $y$ directions. It is  difficult to treat numerically   the dipolar interaction integral in  Eq.  (\ref{GP3d}) 
in configuration space due to the problematic $1/|{\bf R}|^3$ term.
The dipolar interaction integral is evaluated in the momentum space by a Fourier transformation and this is advantageous numerically as that integral in momentum space has a smooth behavior.  The Fourier transformation of the dipolar potential can also be obtained analytically and this aids in the numerical solution of Eq.  (\ref{GP3d}) \cite{dip}. After evaluation in momentum space, the results are transformed back to configuration space by a backward Fourier transformation.

 In numerical calculation, we use
the parameters of $^{164}$Dy atoms, e.g., $a_{\mathrm{dd}} = 130.8a_0$ and $m = 164$ amu. 
The  parameter $K_3$ is not experimentally known but it should have a small effect, compared to the usual two-body term, proportional to  $4\pi a$, in Eq. (\ref{GP3d}). The order of magnitude of the imaginary part of $K_3$ is known \cite{3bdyexp} and we take the real part of $K_3$ in this study to have similar values. 
Here we  consider  $K_3= 10^{-39}$ m$^6$/s and  $K_3= 10^{-38}$ m$^6$/s.  
With  the unit of length $l = 1$ $\mu$m, we have for  unit of time $ ml^2/\hbar = 2.58$ ms, 
for unit of 3D density   \;  $l^{-3}= 1$ $\mu$m$^{-3}$,
for
 unit of energy $\hbar^2/(ml^2)
 = 4.08 \times 10^{-32}$ J, and for  unit of $K_3$ $\hbar l^4/m = 3.87\times 10^{-34}$ m$^6$/s.
To study the effect of a variation of the scattering length $a$ on the formation of a self-bound dipolar BEC, 
we vary the scattering length for values smaller than  its experimental value $a= (92\pm 8)a_0$ \cite{expdy}. In an  experiment, a variation of the value of scattering length is effected by manipulating an external magnetic field near a  Feshbach resonance \cite{feshbach}. 
 For the formation of a dipolar droplet,  we need a strongly dipolar BEC with $\varepsilon_{\mathrm{dd}} >1$ and a BEC of $^{164}$Dy atoms with experimental $\varepsilon_{\mathrm{dd}}\equiv 
a_{\mathrm{dd}}/a =1.4217 >1$ is an ideal candidate for the same.

\begin{figure}[t!]
\begin{center}
\includegraphics[width=.58\linewidth]{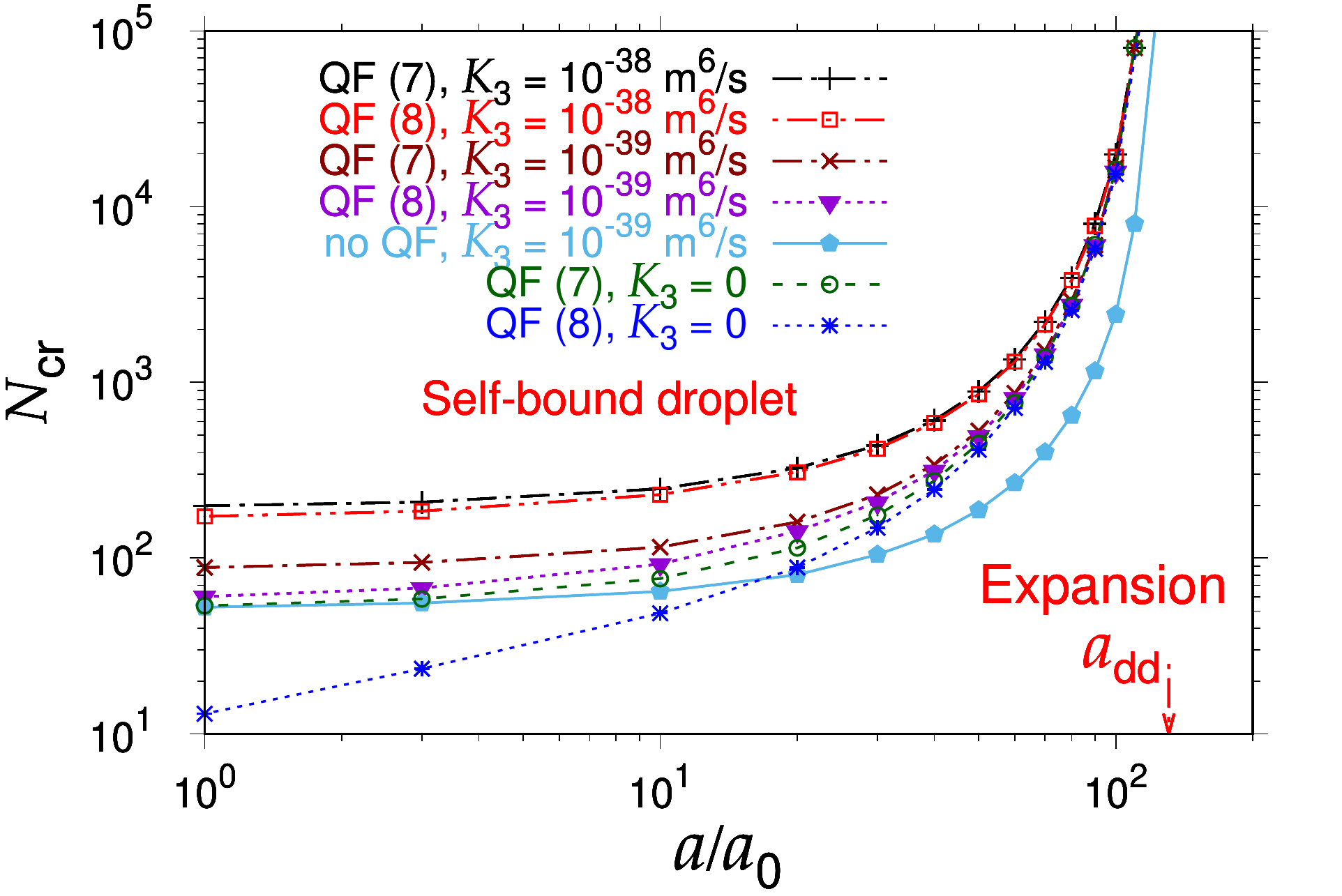}

\caption{Variational critical number of atoms $N_{\mathrm{cr}}$ for the formation of a self-bound dipolar
droplet  for $K_3=0$, $K_3=10^{-39}$ m$^6$/s, and  $K_3=10^{-38}$ m$^6$/s and QF coefficients (\ref{eq7}) and (\ref{eq8}). A result with zero QF coefficient (no QF) is also shown.
For $N > N_{\mathrm{cr}}$ a self-bound dipolar droplet can be formed. For $a> a_{\mathrm{dd}} $ and for $N< N_{\mathrm{cr}}$ there is  expansion to infinity and no droplet can be formed. Plotted quantities are dimensionless.}

\label{fig2} 
\end{center}
\end{figure}

The effect of different beyond-mean-field interactions $-$ the QF LHY interaction (\ref{eq7}), its approximation (\ref{eq8}), and the three-body interaction $-$ on the formation of a dipolar droplet in free space
for different $a$ and $N$  can be  qualitatively understood from a consideration of the analytic variational energy (\ref{energy}). In Eq. (\ref{energy}), all terms are positive (repulsive) except the contact and dipolar interaction terms involving $a$ and $a_{\mathrm{dd}}$, respectively.  For the formation of a self-bound droplet, the energy (\ref{energy}) has to be negative for certain $w_\rho$ and $w_z$,  and that happens for  $N$  larger than a critical value  $N_{\mathrm{cr}}$, while Eqs. (\ref{w1}) and (\ref{w2}), for variational widths $w_\rho$ and $w_z$, allow  real solutions.  For $N < N_{\mathrm{cr}}$ the system is
much too repulsive and undergoes an expansion  to infinity without the formation of a droplet. However, the critical number  $N_{\mathrm{cr}}$  is a
function of the three-body coefficient $K_3$ and scattering length $a$.  The scattering length $a$ can be controlled experimentally, independent of the three-body  coefficient  $K_3$, by using an  optical \cite{32} or a magnetic \cite{feshbach}  Feshbach resonance. There are also different suggestions for controlling $K_3$ similarly \cite{K3vary}.  
As $a$ increases, the system becomes less attractive  and it is possible to have a negative energy only  for a larger  $N$. Consequently, $N_{\mathrm{cr}}$ increases as $a$ increases for a fixed $K_3$. Similarly, a non-zero $K_3$ makes the system less attractive and a larger  $N$ is needed to make the energy negative. Hence $N_{\mathrm{cr}}$ should increase monotonically  with $K_3$, while $a$ is held fixed.  
 The $N_{\mathrm{cr}}$-$a$ correlation
for  $K_3=0$,  $ = 10^{-39}$ m$^6$/s, and  $ = 10^{-38}$ m$^6$/s,   is shown in Fig.  \ref{fig2} for QF LHY interaction (\ref{eq7}) and its often used \cite{blakie1,blakie2,blakie,pohl,pfau} approximation (\ref{eq8}).
To demonstrate that the three-body repulsion alone can form a self-bound droplet, we also display in Fig. \ref{fig2} the result for 
$K_3 = 10^{-39}$ m$^6$/s with no LHY interaction. 
 For small  $K_3$ ($K_3=0$ and $K_3=10^{-39}$ m$^6$/s) and small $a$, the $N_{\mathrm{cr}}$-$a$ correlation of Fig. \ref{fig2} is sensitive to the form of QF coefficient $-$ Eq. (\ref{eq7}) or Eq. (\ref{eq8}). For a large $K_3$ ($K_3=10^{-38}$ m$^6$/s), 
the associated strong three-body repulsion dominates at short distances, that musks the effect of QF LHY interaction,
and the result is not sensitive to the form of $\gamma_{\mathrm{QF}}$.  The result with  zero QF LHY interaction and $K_3=10^{-39}$ m$^6$/s is quite similar to  the result with QF LHY interaction
 (\ref{eq7})  and $K_3=0$, specially for small $a$; in that case the use of $K_3$ alone is recommended  for the study of dipolar droplets.

\begin{figure}[t!]
\begin{center}
\includegraphics[width=.25\linewidth]{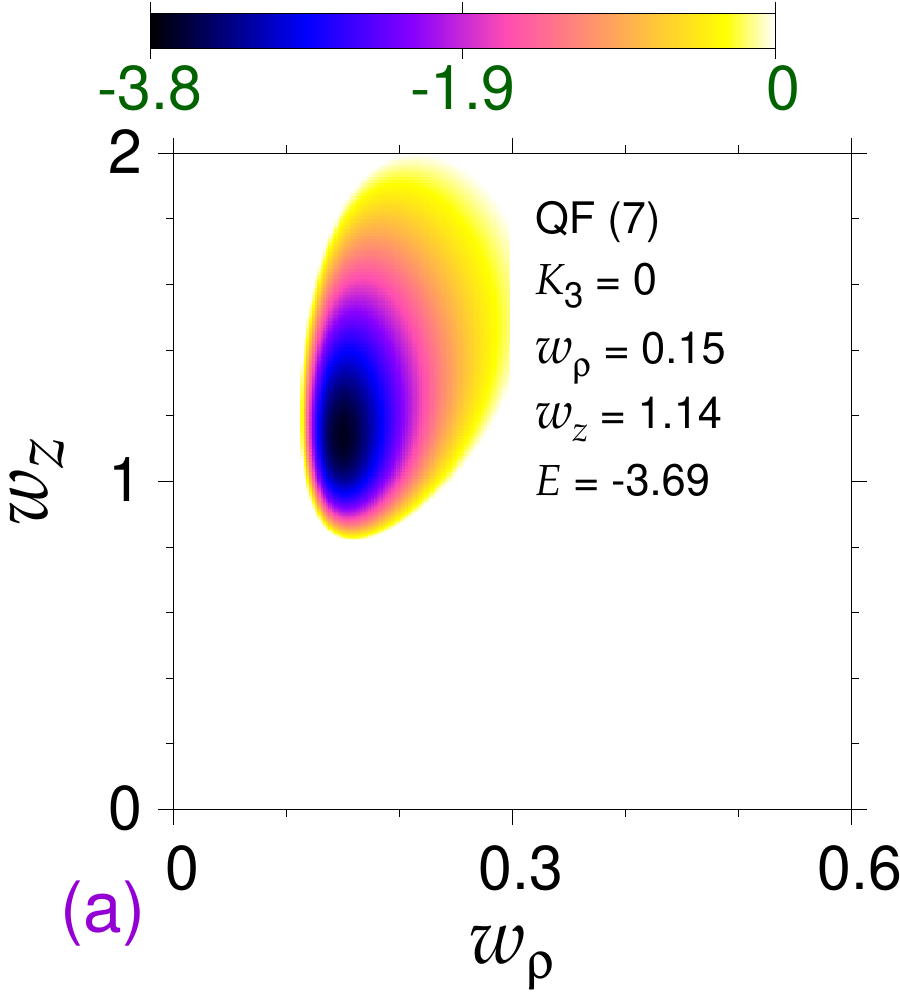}
 \includegraphics[width=.25\linewidth]{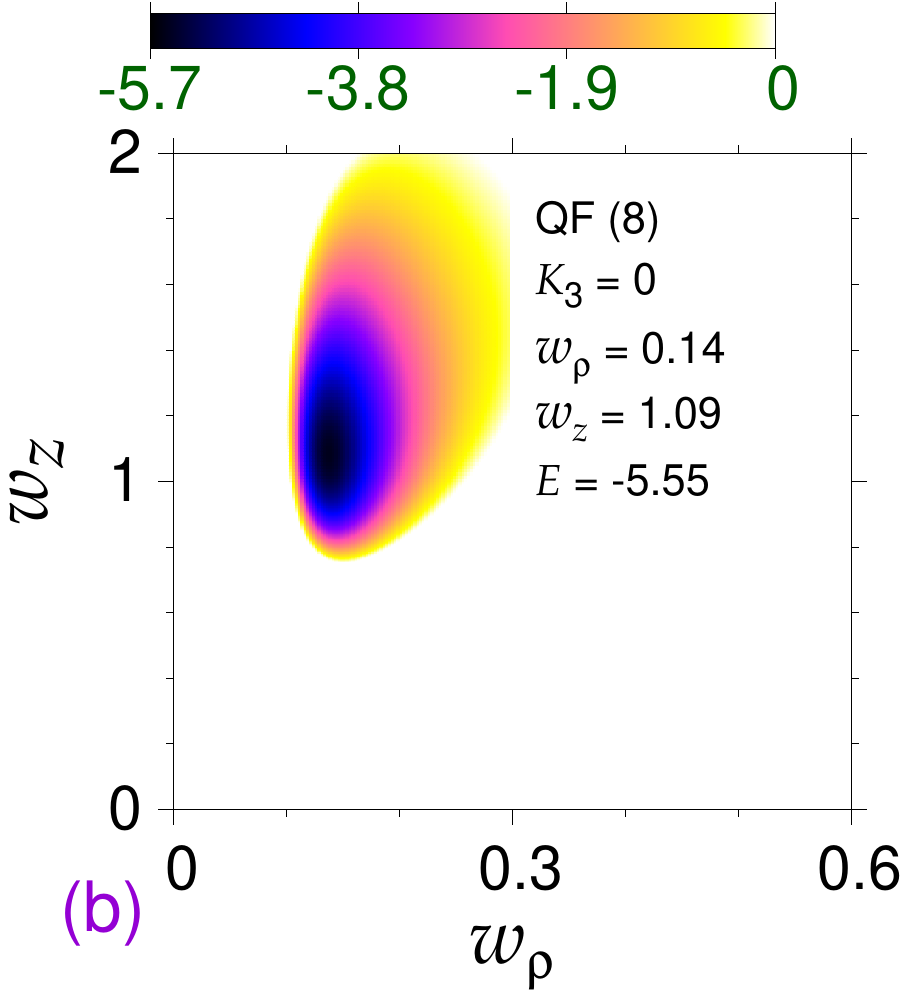}
 \includegraphics[width=.25\linewidth]{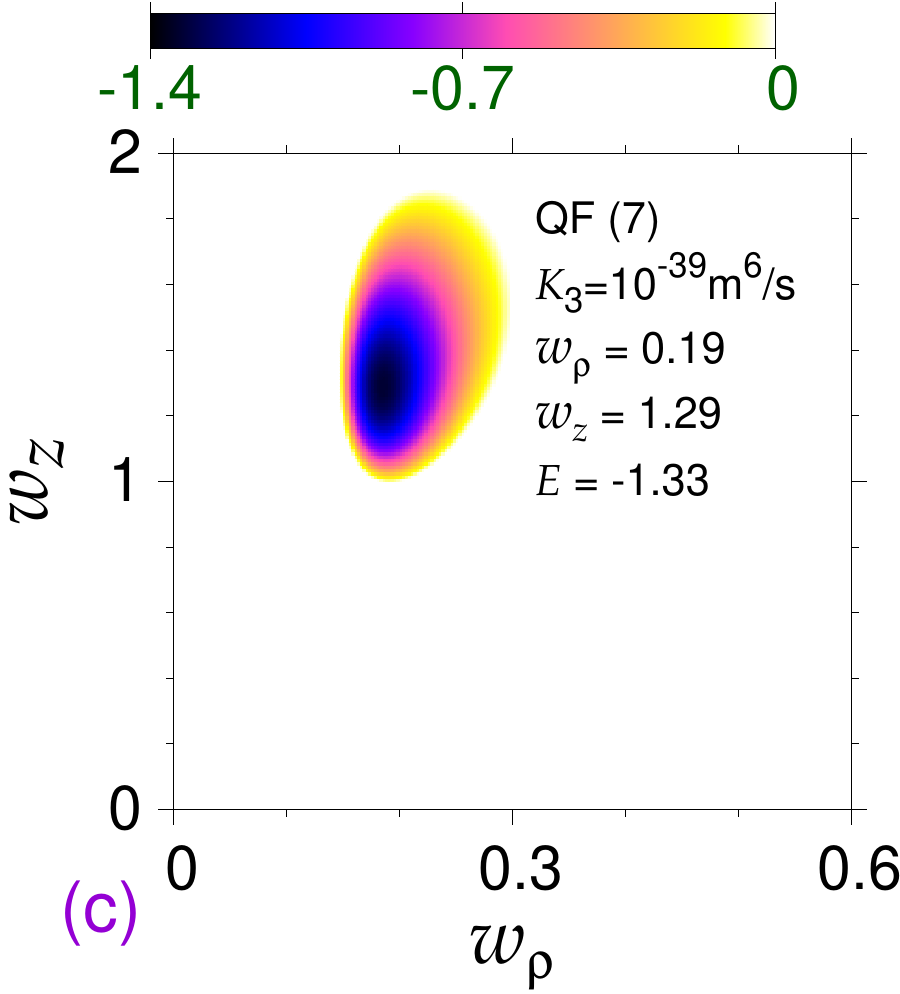}

\caption{ 2D contour plot of variational energy per atom (\ref{energy}) highlighting its  minimum value  and
the negative energy region for  $a=60a_0, N=1000$  $^{164}$Dy atoms, as a function of widths $w_\rho$ and $w_z$ for
(a) $K_3=0$, QF coefficient  (\ref{eq7}), (b) $K_3=0$, QF coefficient  (\ref{eq8}), and (c)   $K_3=10^{-39}$ m$^6$/s,   QF coefficient (\ref{eq7}). Plotted quantities   are
dimensionless and the physical unit for $^{164}$Dy atoms can be restored using the unit
of length $l = 1$ $\mu$m.
 }
\label{fig3} 
\end{center}
\end{figure}

\begin{figure}[b!]
\begin{center}
\includegraphics[width=.55\linewidth]{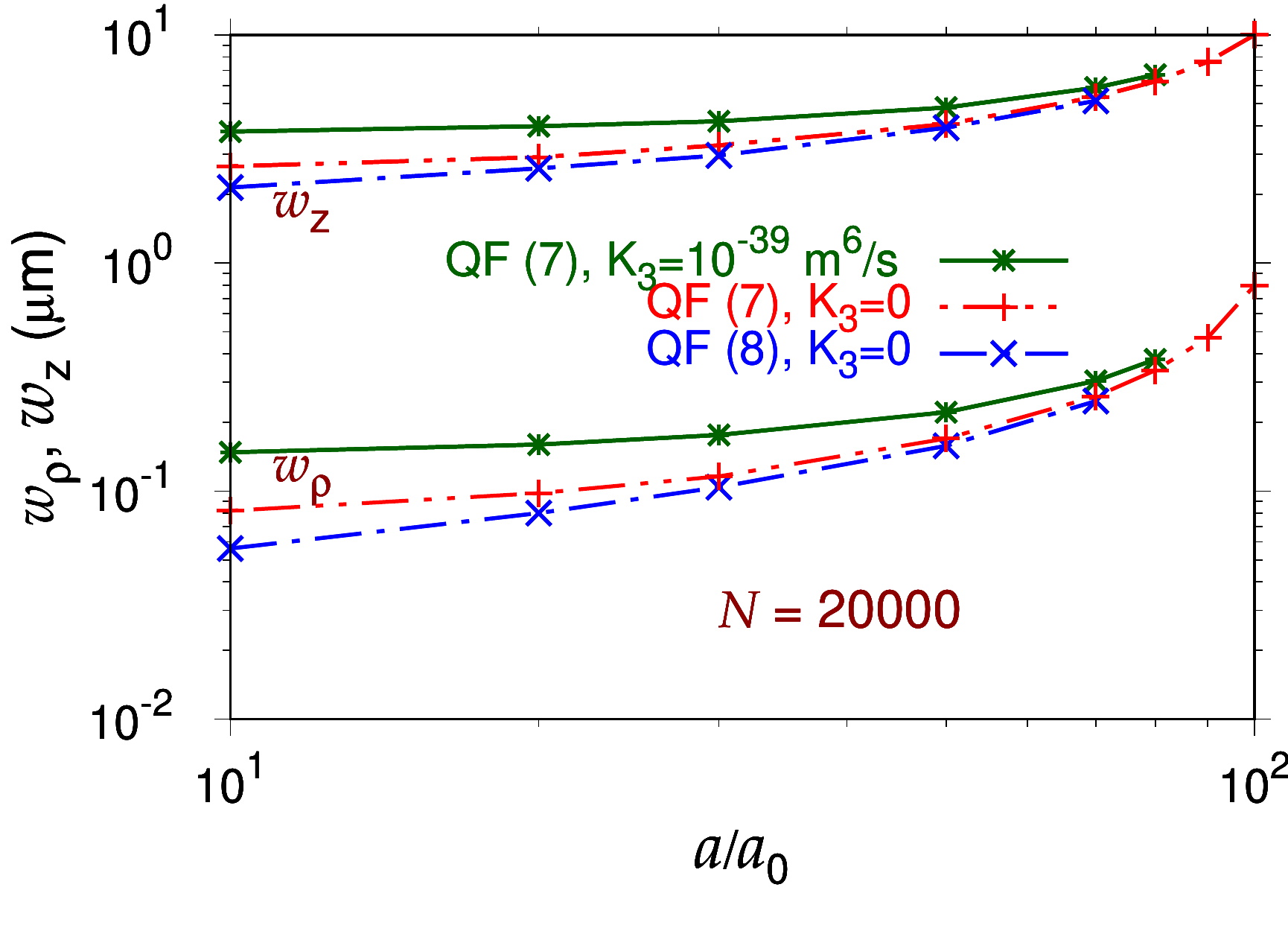}

\caption{Variational results  for  widths $w_\rho$ and $w_z$
 for different beyond-mean-field interaction. The lines are labeled by the   QF coefficients (\ref{eq7}) and (\ref{eq8}) and the values of the three-body coefficient $K_3$.}
\label{fig4} 
\end{center}
\end{figure}

In Fig.  \ref{fig3} we display the 2D contour plot of variational energy (\ref{energy}) as a function
of widths $w_\rho$ and $w_z$ for different  beyond-mean-field interactions and $N=1000, a=60a_0$.
The plots in this  figure highlight the negative energy
region for different $w_\rho$ and $w_z$. The minimum negative energy in these plots corresponds to 
a self-bound dipolar droplet. 
 The white region in these plots corresponds to positive energy.  If we compare plots \ref{fig3}(a) and (b)
 we find that an increased attraction  due to  the use of approximate QF coefficient  has reduced  the energy by about $50\%$.
Similarly, comparing plots \ref{fig3}(a) and (c), we find that the inclusion of a moderately repulsive three-body interaction has increased  the energy by about $60\%$.     In both cases there has been a change in the profile of the self-bound state with a change in the widths $w_\rho$ and $w_z$ due to  the use of different beyond-mean-field interactions (result not presented in this paper).

\begin{table}
\label{TI}
\caption{Variation and numerical energy $E$,  rms size $\langle x\rangle\equiv w_\rho/\sqrt 2$  and  $\langle z\rangle\equiv w_z/\sqrt2$ of  self-bound   axially-symmetric dipolar    droplets of $^{164}$Dy atoms in dimensionless units  for different $a$, $N$, $\gamma_{\mathrm{QF}}$ and $K_3$.}
 \begin{tabular}{|l|l|l|l|c|c|c|c|c|l|}\hline
$a$&$N$&$\gamma_{\mathrm{QF}}$ &$K_3$ &\multicolumn{3}{c|}{Variational} & \multicolumn{3}{c|}{Numerical}\\
 $(a_0)$       &(10$^3$)& &(m$^6$/s)  & $\langle x \rangle$  & $\langle z \rangle$& $E$  & $\langle x \rangle$  & $\langle z \rangle$ & $E$  \\
\hline
90 & 10&(\ref{eq7})  &0& 0.32&  4.02 & $-0.74$& 0.34 &3.83  &-0.95  \\
90 & 10& (\ref{eq7})& 10$^{-39}$&0.35 & 4.20&$-0.59$ & 0.36 &4.01  & -0.76 \\
90 & 10&(\ref{eq7}) &10$^{-38}$&0.50 & 5.22 & $-0.14$ &0.53  &5.00  &-0.23 \\ \hline
90 & 10& (\ref{eq8}) &0& 0.31&  3.89 & $-0.93$& 0.32& 3.70 & $-1.15$ \\
90 & 10& (\ref{eq8}) & 10$^{-39}$&0.33 & 4.09 &$-0.72$ & 0.35 & 3.89 & $-0.92$ \\
90 & 10& (\ref{eq8}) &10$^{-38}$&0.49 & 5.12 & $-0.17$ & 0.51 & 4.77 & $-0.26$\\ \hline
30 &0.3 &  (\ref{eq7})&0  & 0.042 & 0.29 & $-46.3$ &0.045 & 0.27 & $-56.2$ \\
30 &0.3 & (\ref{eq7}) & 10$^{-39}$&0.065 & 0.37 & $-10.0$ & 0.070 & 0.35 & $-13.8$ \\
30 &0.3 &  none &10$^{-39}$&0.038 & 0.27 & $-130$ & 0.038 & 0.25 & $-141$\\
 \hline
30 & 0.3 & (\ref{eq8}) &0 &  0.037 & 0.26&$-53.9$  & 0.037 & 0.27 & $-69.1$ \\
30 & 0.3& (\ref{eq8}) &10$^{-39}$&0.061 & 0.35 & $-16.2$ & 0.063 & 0.33 & $-21.6$ \\
 \hline
 
\end{tabular}
\end{table}

 To demonstrate that the formation of a self-bound droplet  remains sensitive to the use of different 
beyond-mean-field interaction for a large number of atoms $N$, specially for small scattering lengths $a$, we illustrate in Fig. \ref{fig4} the variational widths $w_\rho$ and $w_z$ of a self-bound droplet of $N=20000$ atoms versus $a$ for different beyond-mean-field interaction controlled by the QF coefficients (\ref{eq7}) and (\ref{eq8}), and three-body coefficient $K_3$. We find that the widths can have a large variation
for the use of different beyond-mean-field interaction for small values of $a$.  In Fig. \ref{fig1}, we find that the QF coefficient  (\ref{eq7}) is more repulsive than its approximation (\ref{eq8}). Consequently, for $K_3=0$, in Fig. \ref{fig4} the widths obtained with the use of QF  coefficient (\ref{eq7}) 
are larger than the same obtained with the use of Eq. (\ref{eq8}). The inclusion of a non-zero three-body coefficient $K_3$ increases the repulsion and consequently, the widths increase. Hence from Figs. \ref{fig3} and \ref{fig4} we find that the formation of a self-bound droplet $-$ its shape and size $-$ is sensitive to the use of different beyond-mean-field interaction.

In Table I we compare the variational and numerical results for the rms sizes $\langle x \rangle$, and $\langle z \rangle $ and energy per atom $E$ of a self-bound dipolar droplet for different $a$, $N$, $\gamma_{\mathrm{QF}}$, and $K_3$. The agreement between the variational and numerical results is satisfactory.
The dipolar system is more attractive for $a=30a_0$ than for $a=90a_0$ due to an increased  contact repulsion in the latter, thus resulting in strongly-bound droplets for $a=30a_0$  with  smaller size  $\langle z \rangle$  and  smaller energy. 
The attraction also increases with a reduction in $K_3$ or with the use of the approximation (\ref{eq8}) in place of the exact QF coefficient (\ref{eq7}), leading to smaller energies  and sizes.

\begin{figure}[t!]
\begin{center}
\includegraphics[width=.37\linewidth]{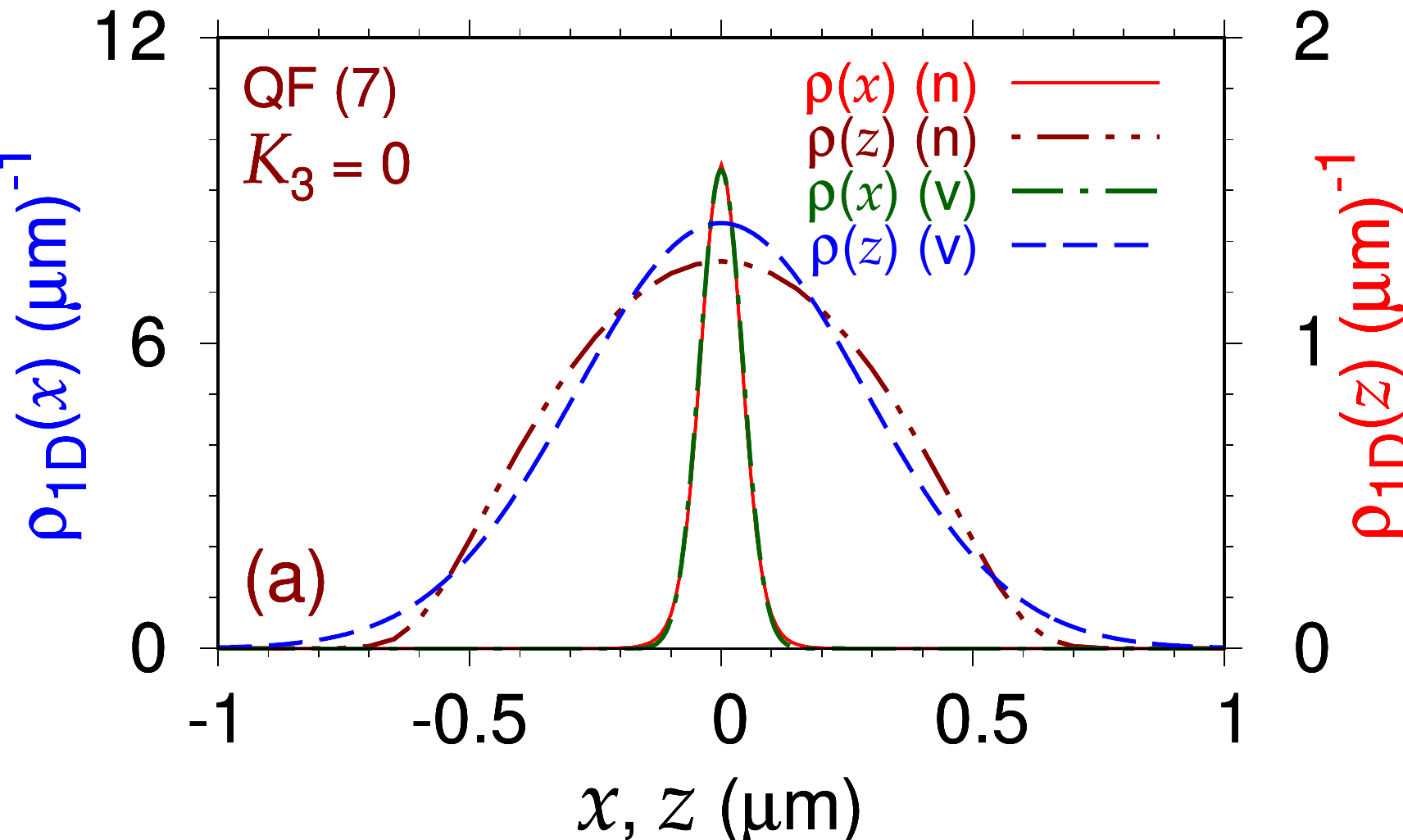}
\hskip .5cm
\includegraphics[width=.37\linewidth]{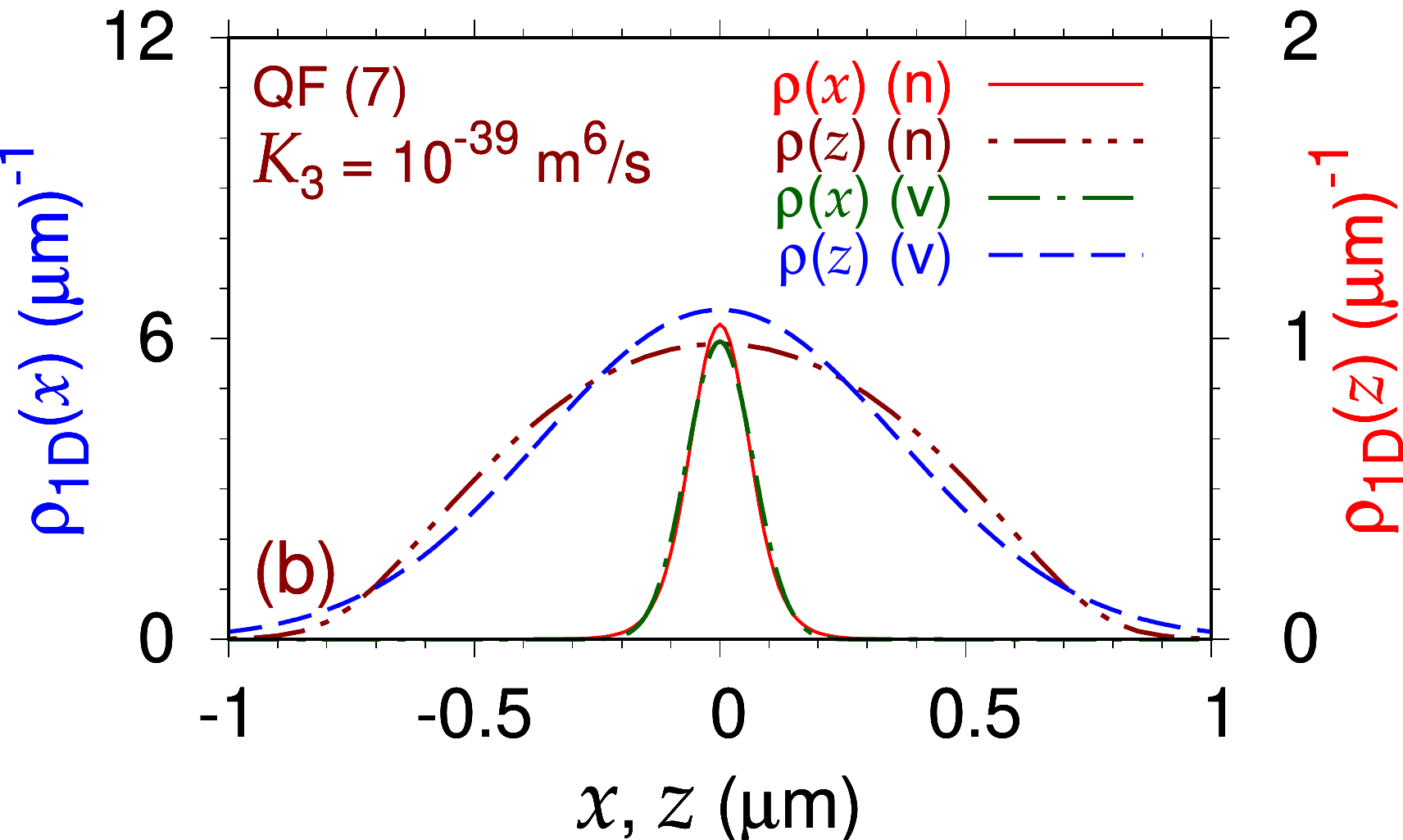}
\includegraphics[width=.37\linewidth]{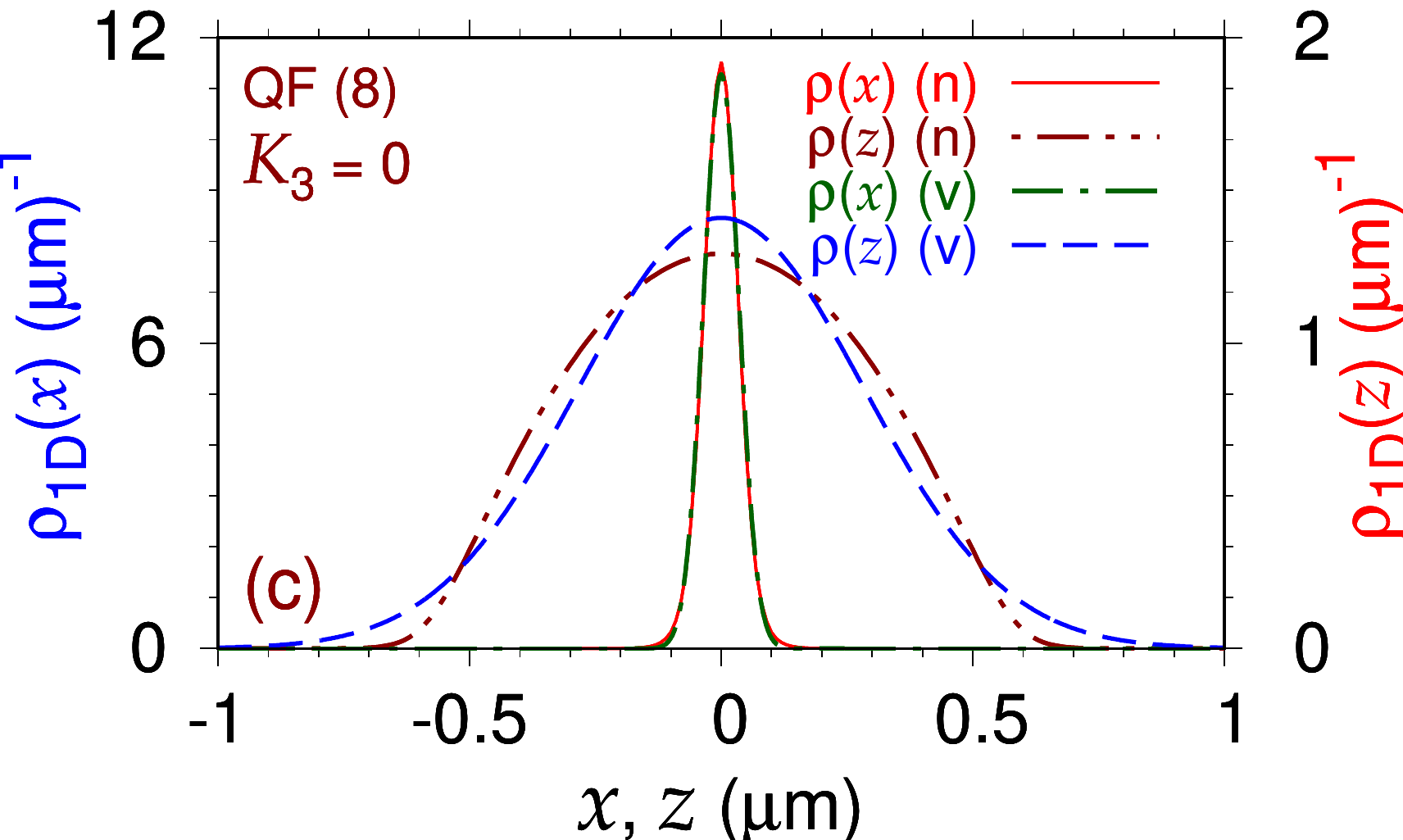}
\hskip .5cm
\includegraphics[width=.37\linewidth]{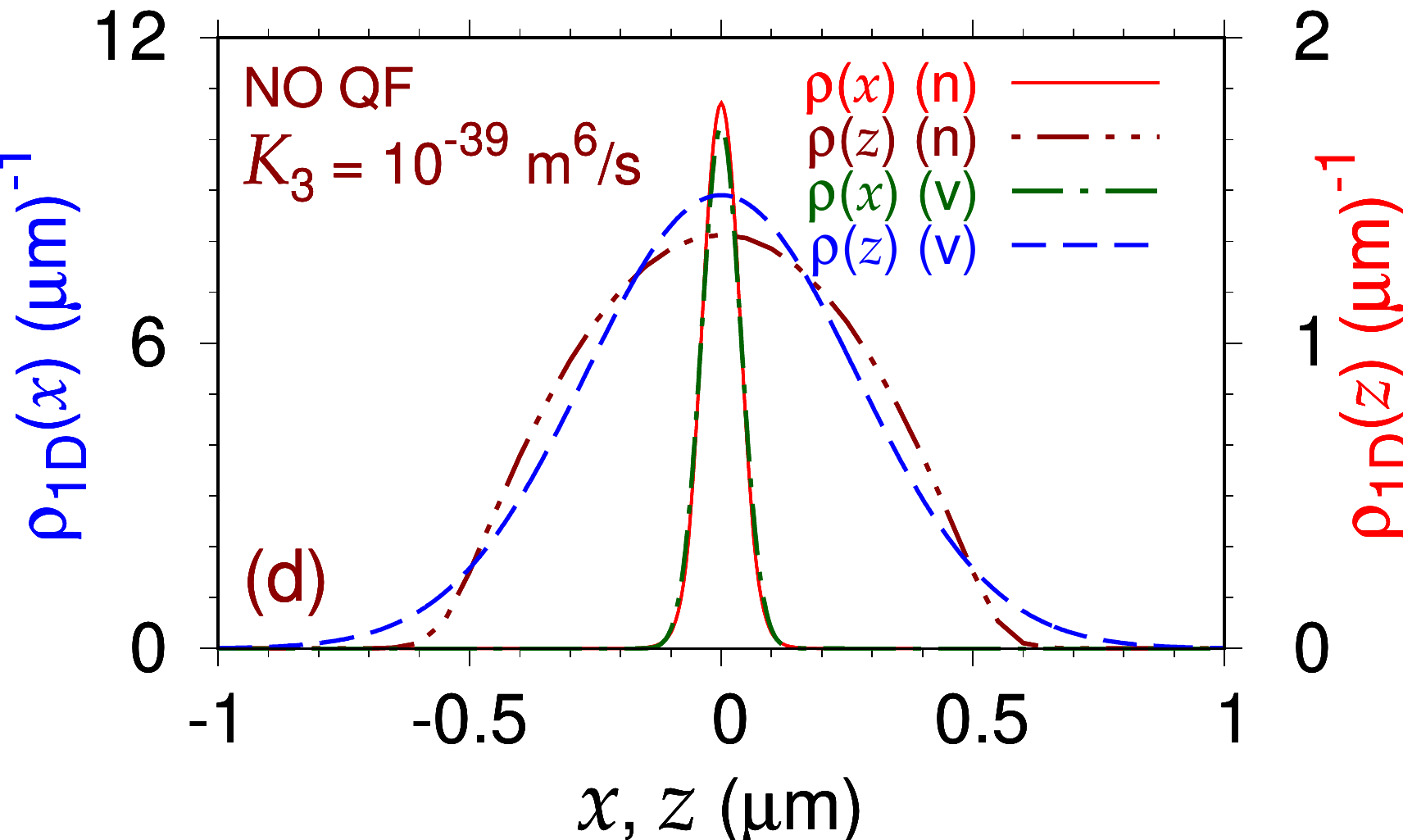}

\caption{  Numerical (n) and variational (v)  reduced 1D
densities $\rho_{1D}(x)$ and $\rho_{1D}(z)$ along $x$ and $z$ directions, respectively,  of a $^{164}$Dy droplet with   $a=30a_0, N = 300,$ and (a) $K_3=0$,  QF coefficient  (\ref{eq7}),
(b) $K_3=10^{-39}$ m$^6$/s,  QF coefficient (\ref{eq7}),
(c)   $K_3=0$,  QF coefficient  (\ref{eq8}), and 
 (d) $K_3=10^{-39}$ m$^6$/s, zero QF coefficient.
}
\label{fig5} 
\end{center}
\end{figure}

\begin{figure}[t!]
\begin{center}
\includegraphics[width=.32\linewidth]{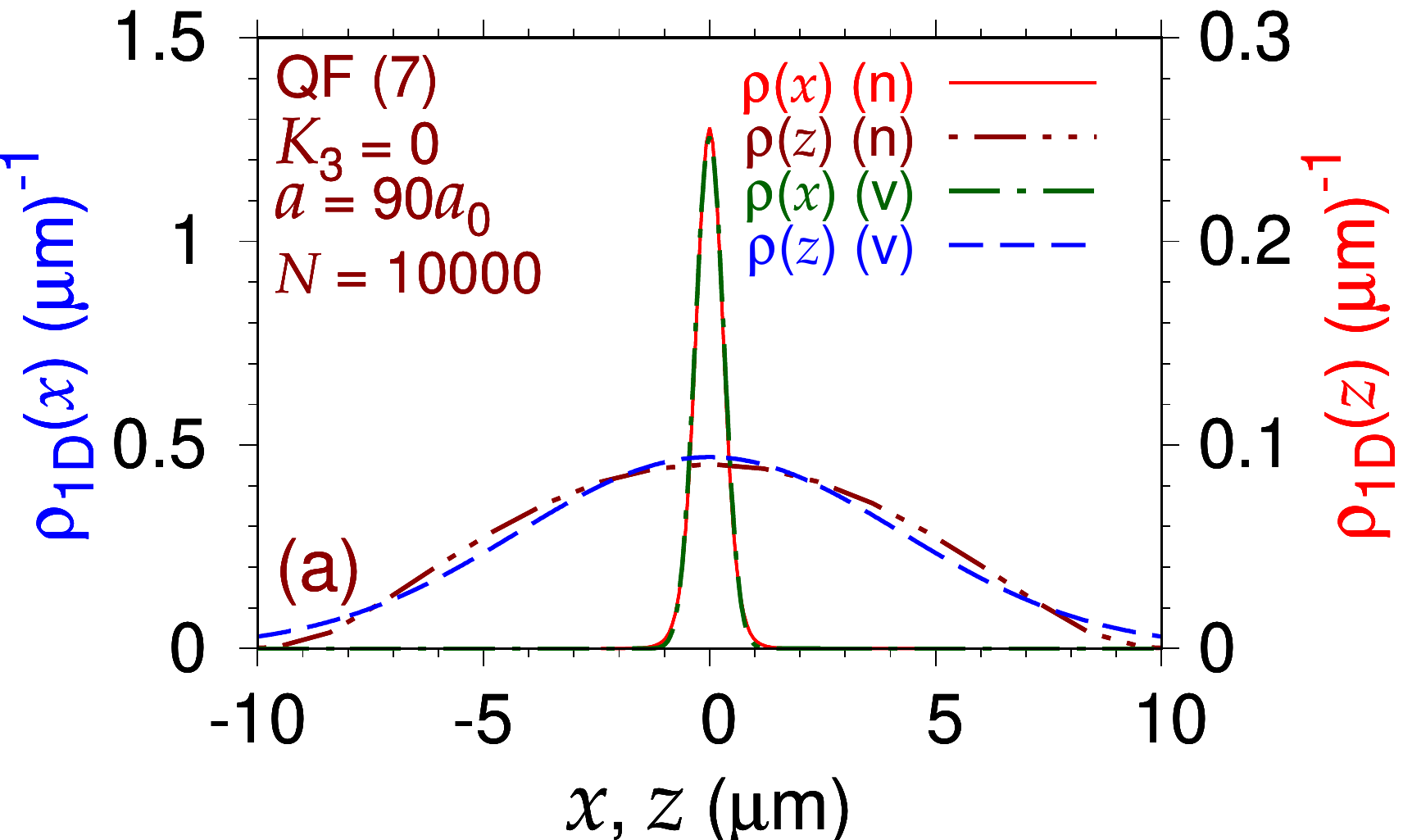}
\hskip .15cm
\includegraphics[width=.32\linewidth]{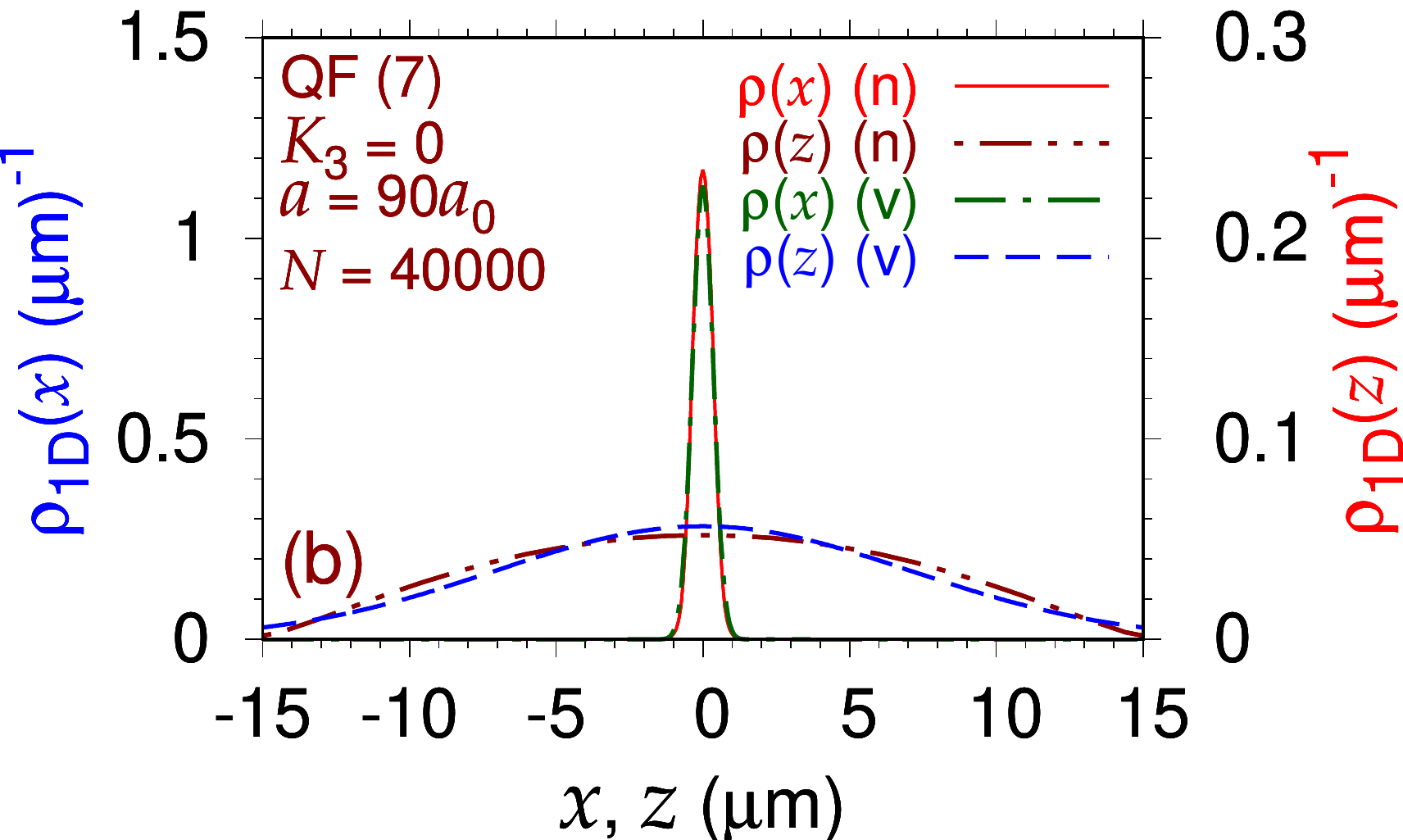}
\hskip .15cm
\includegraphics[width=.32\linewidth]{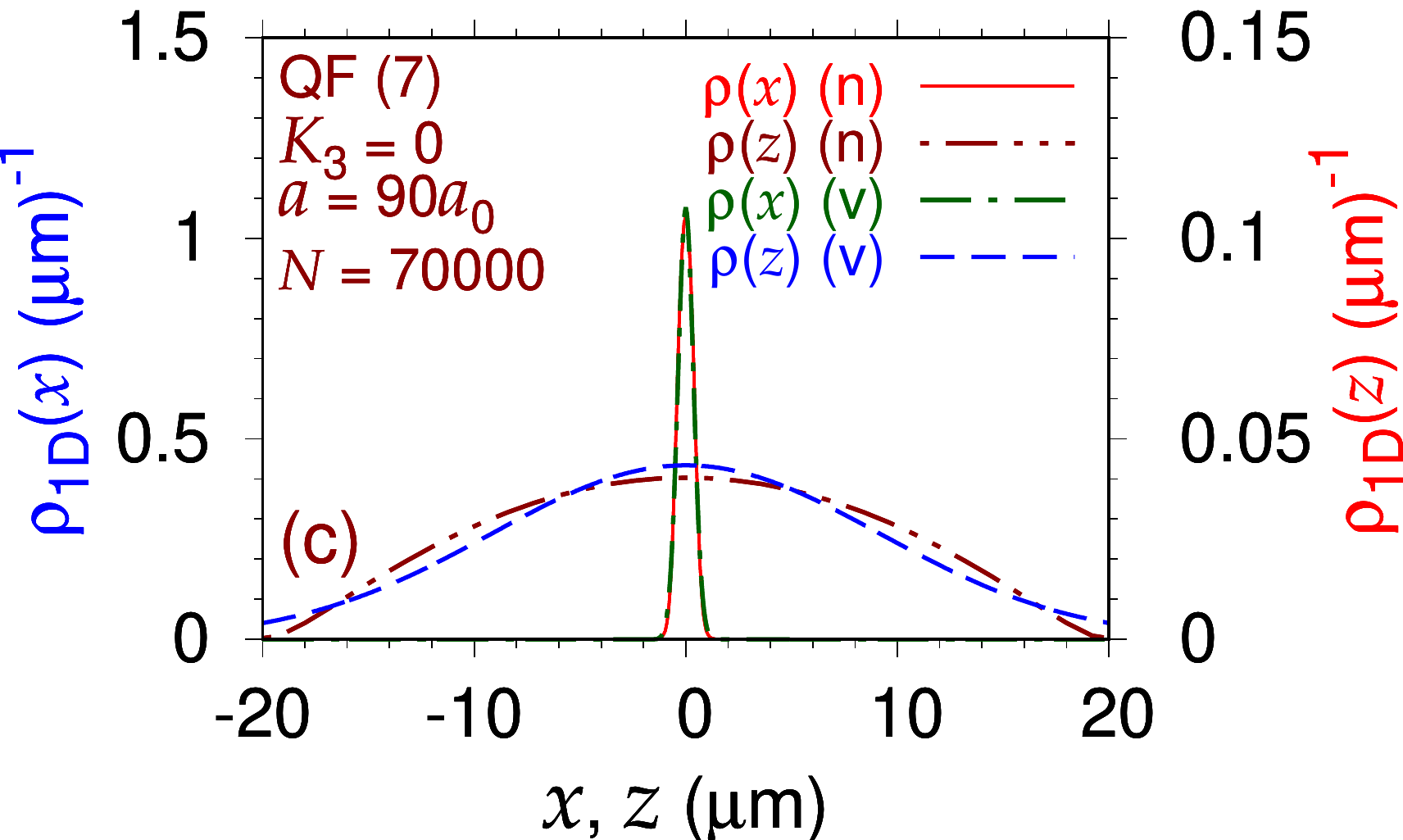}
\includegraphics[width=.32\linewidth]{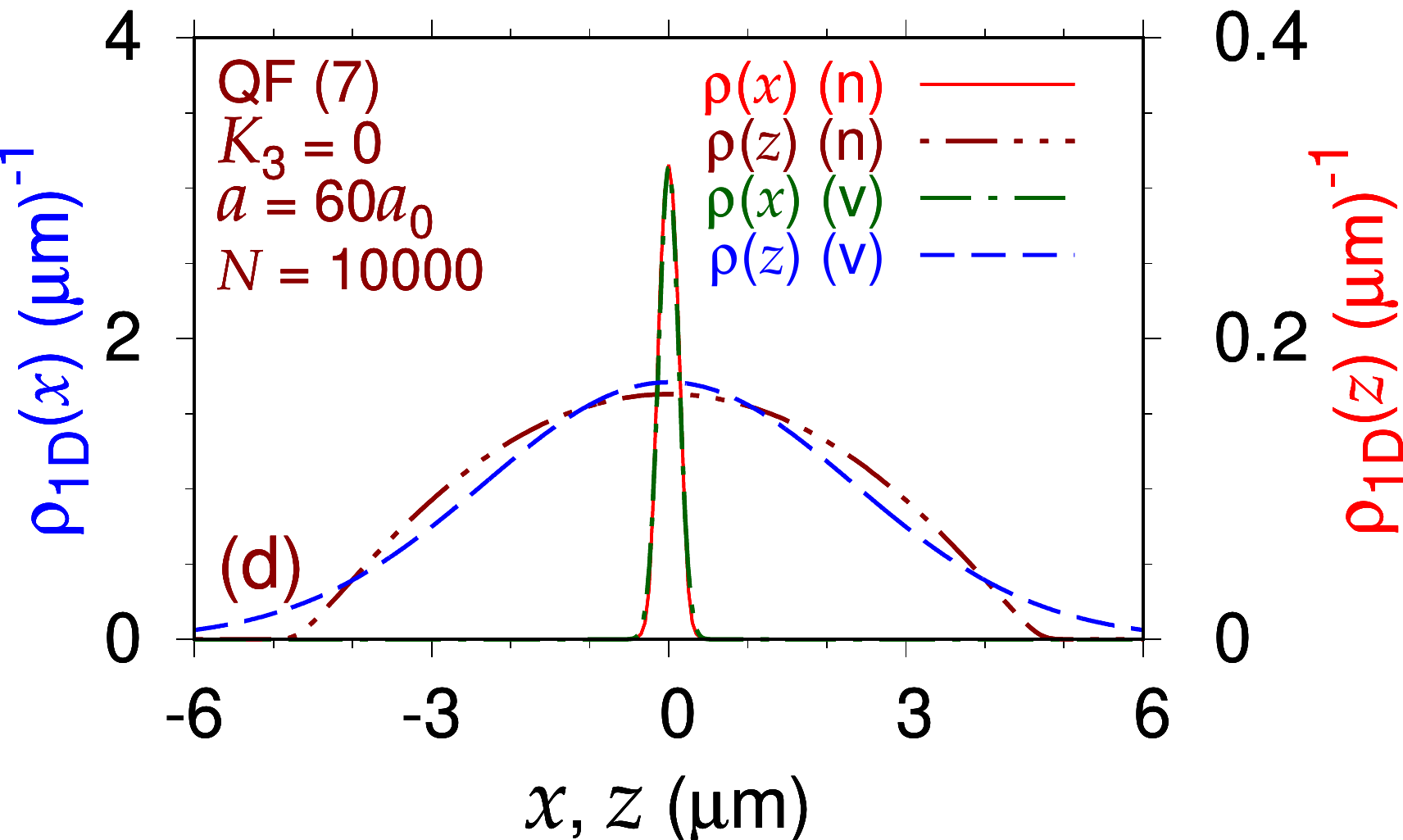} 
\hskip .15cm
\includegraphics[width=.32\linewidth]{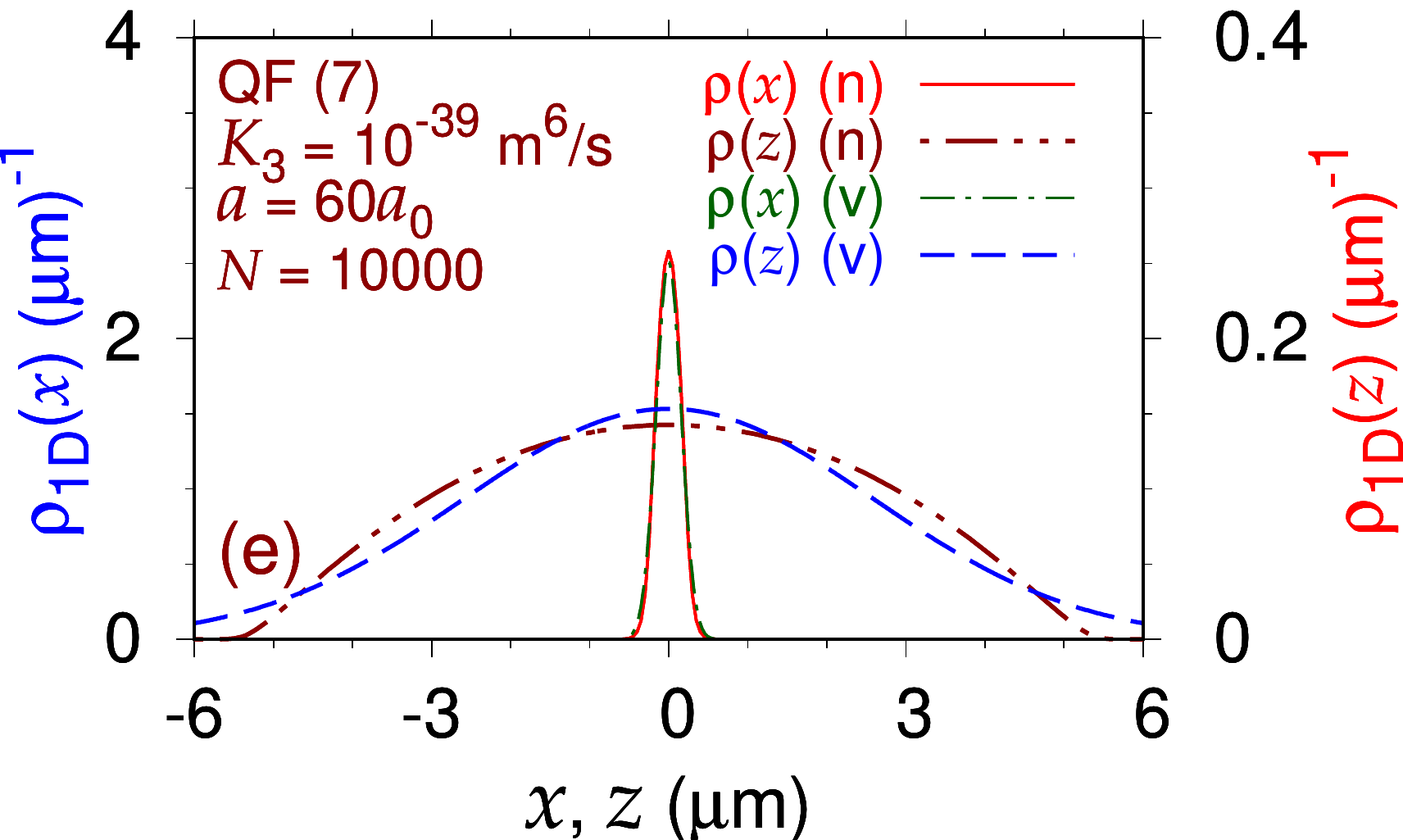}
\hskip .15cm
\includegraphics[width=.32\linewidth]{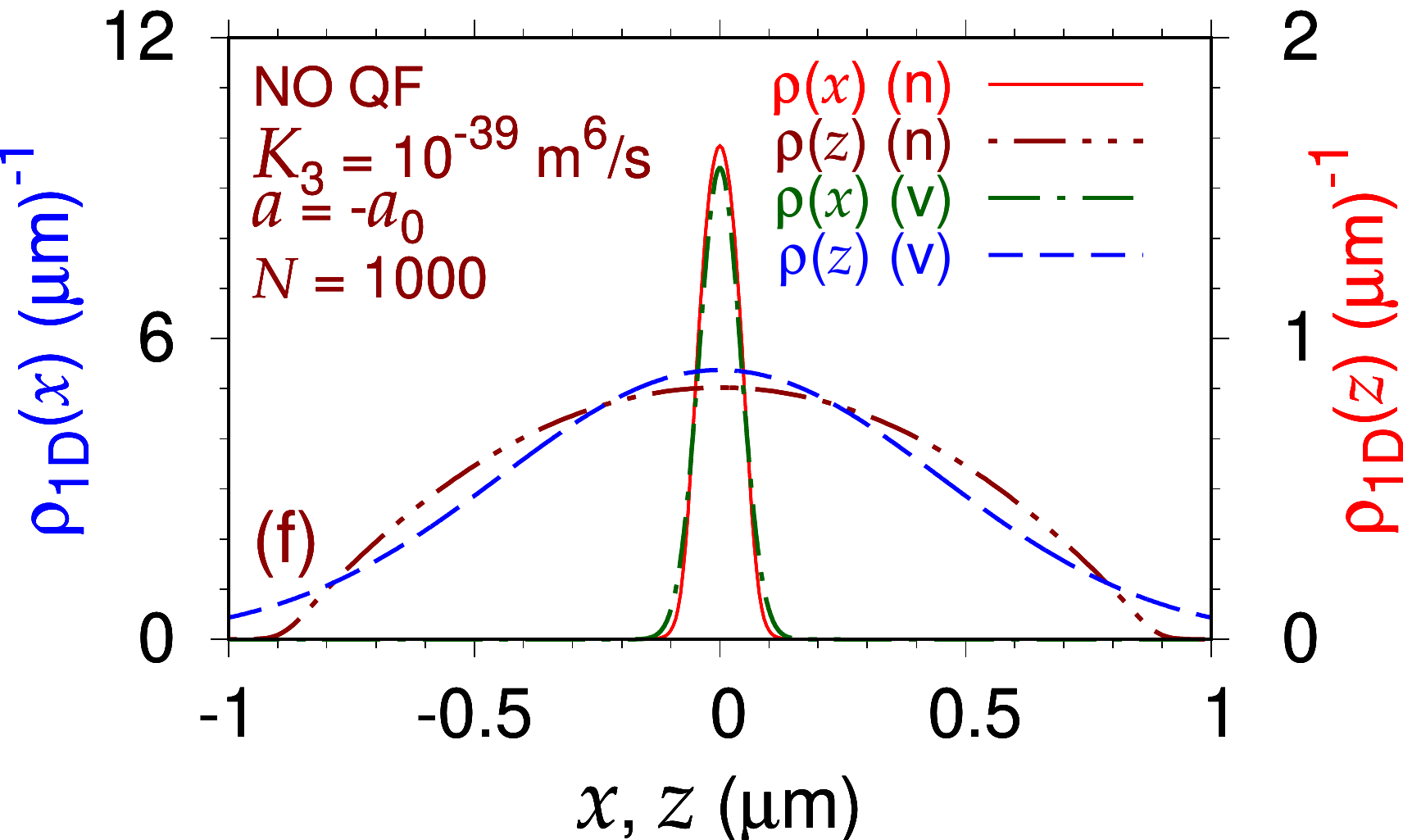} 
  \includegraphics[width=.32\linewidth]{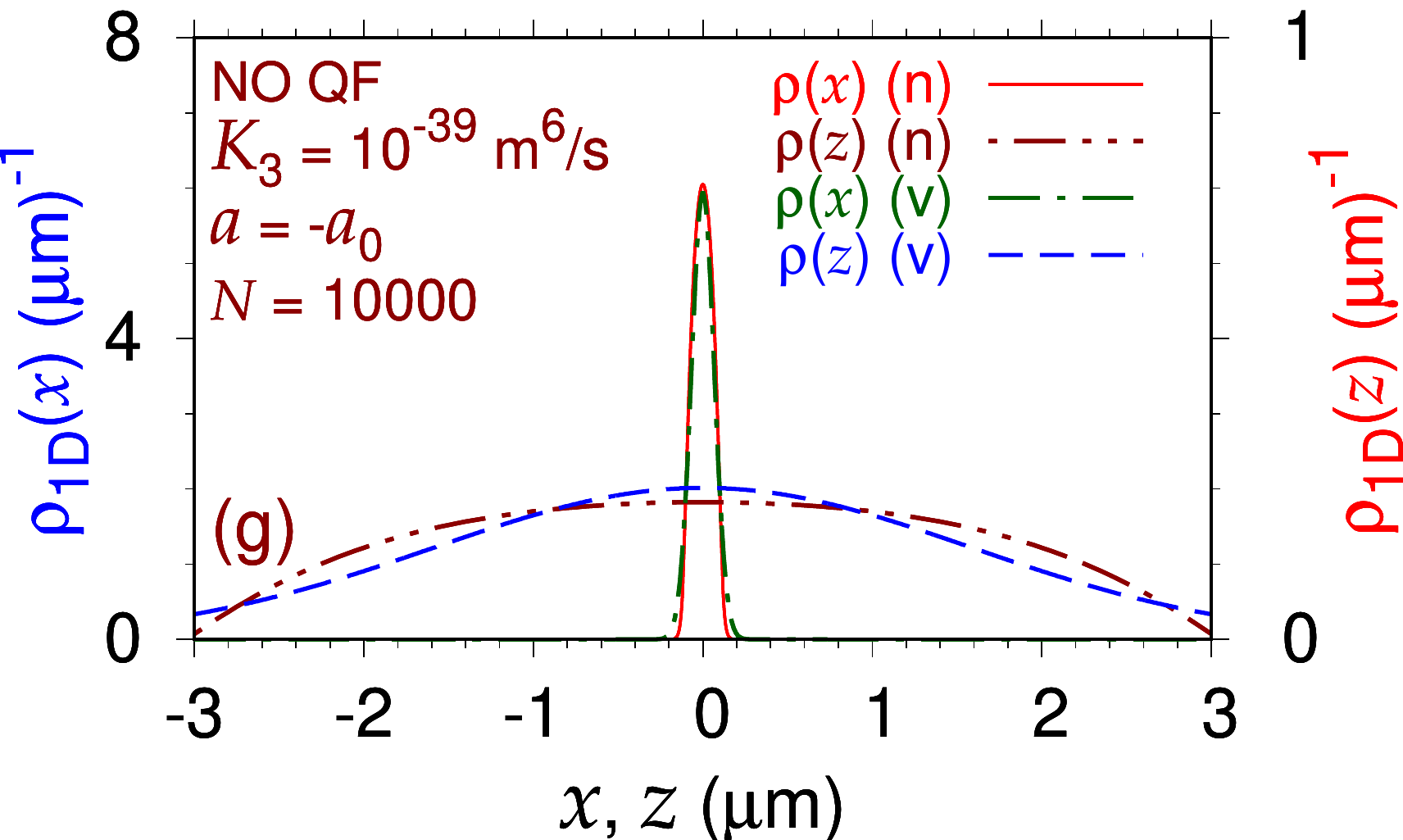}
\hskip .15cm
\includegraphics[width=.32\linewidth]{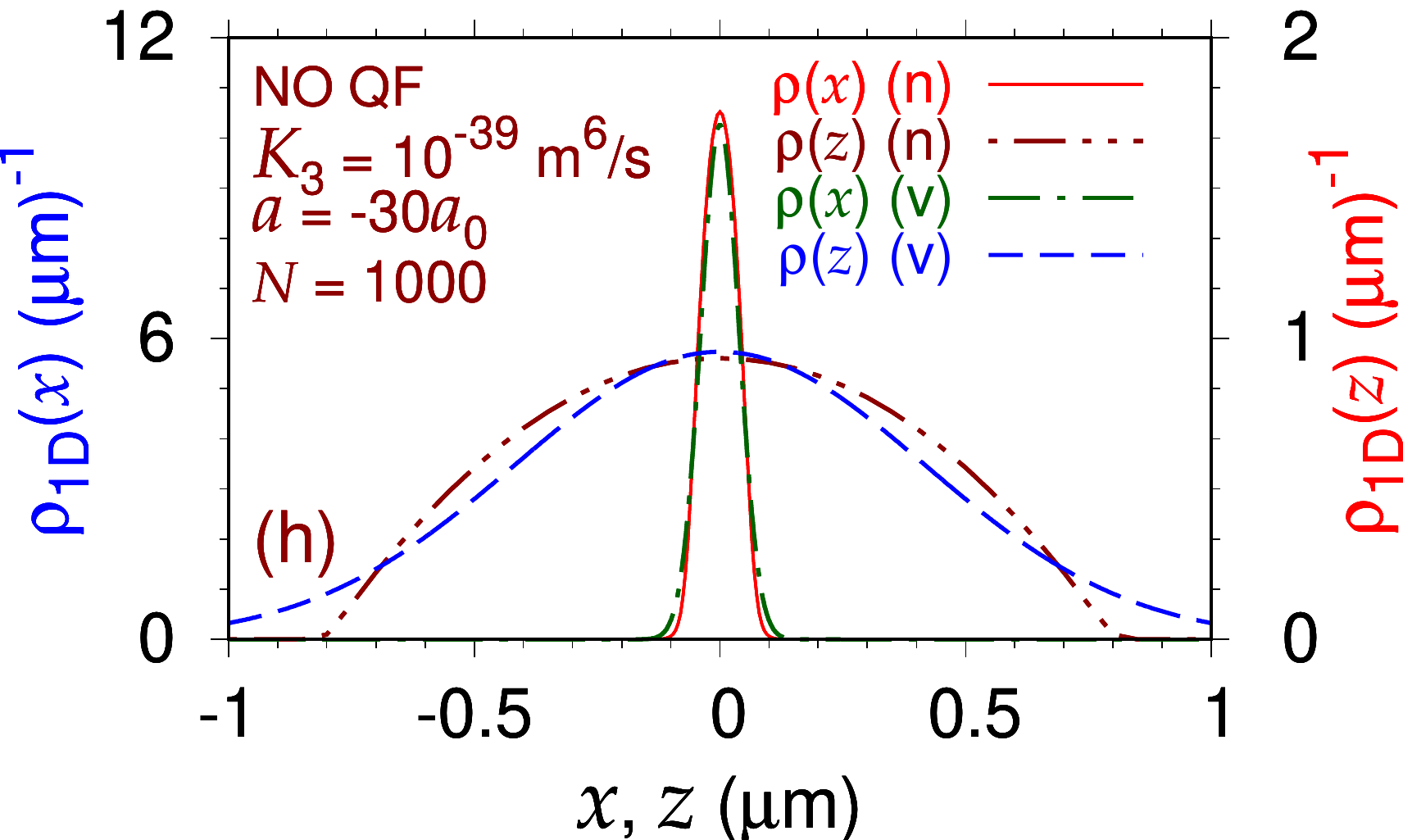}
\hskip .15cm
\includegraphics[width=.32\linewidth]{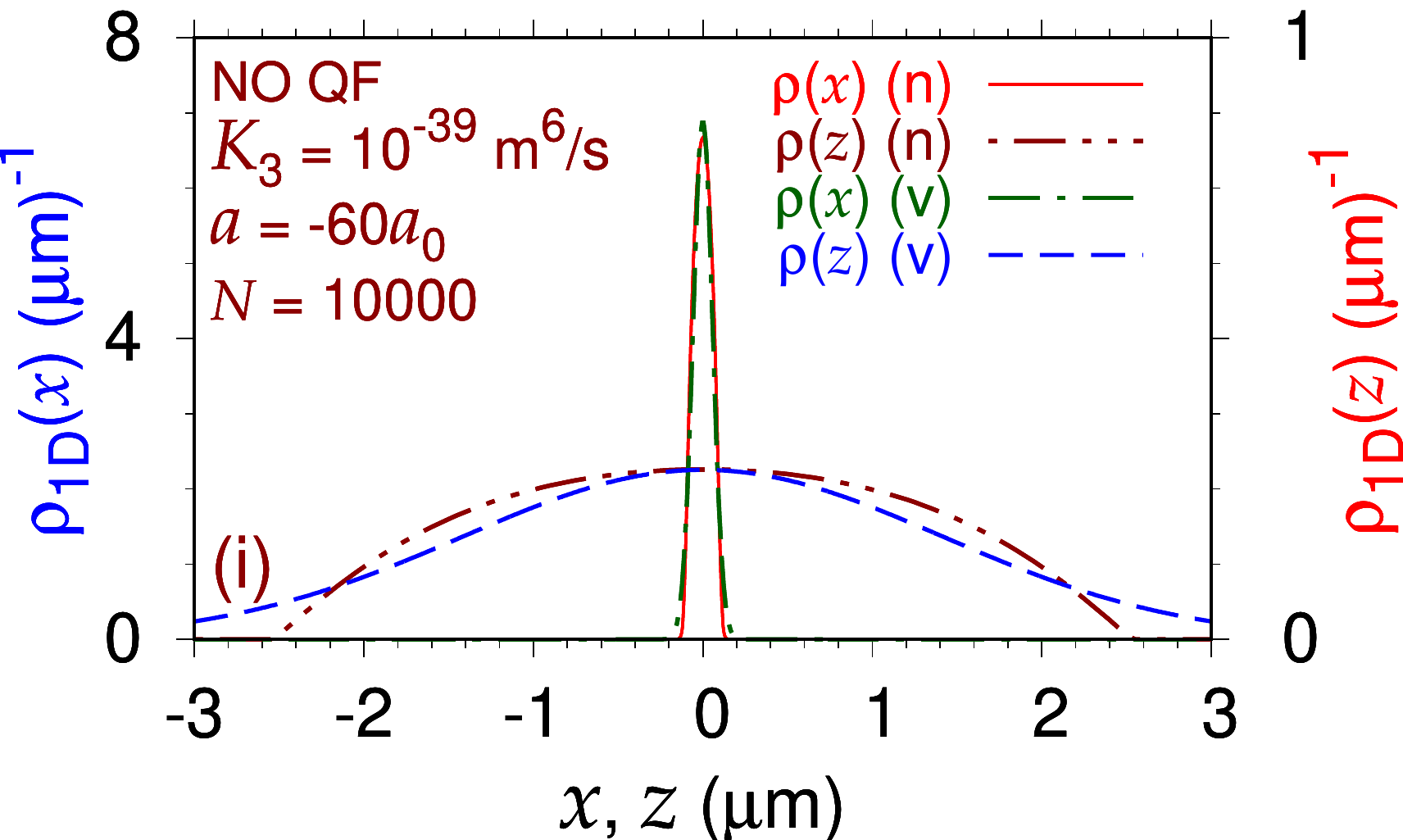}
 
\caption{ Numerical (n) and variational (v) reduced 1D
densities $\rho_{1D}(x)$ and $\rho_{1D}(z)$ along $x$ and $z$ directions, respectively,  of a $^{164}$Dy droplet  with  (a) $a=90a_0, N = 10000,$  (b)  $a=90a_0, N = 40000,$     (c)  $a=90a_0, N = 70000,$   
(d)  $a=60a_0, N = 10000,$ (e)  $a=60a_0, N = 10000,$     (f) $a=-a_0, N = 1000,$  (g)  $a=-a_0, N = 10000,$     (h)  $a=-30a_0, N = 1000,$    and  
(i)  $a=-60a_0, N = 10000.$ In (a)-(e) QF coefficient (\ref{eq7}) is used and in (f)-(i) there is no QF correction. In (a)-(d) $K_3=0$ and $K_3=10^{-39}$ m$^6$/s in others.}
\label{fig6} 
\end{center}
\end{figure}

To study the density distribution of a self-bound   droplet,   we calculate the reduced 1D
densities 
\begin{align}\label{1dx}
\rho_{1D}(x) \equiv \int
dzdy|\psi({\bf r})|^2,  \\ \rho_{1D}(z) \equiv \int 
dxdy|\psi({\bf r})|^2. \label{1dz}
\end{align}
In Fig. \ref{fig5}, we plot these densities as obtained from the variational approximation  and numerical calculation for  $a=30a_0, N=300$, and for different beyond-mean-field interaction, e.g., (a) $K_3=0$ and QF  
coefficient (\ref{eq7}), (b) $K_3=10^{-39}$ m$^6$/s and QF  
coefficient (\ref{eq7}), (c) $K_3=0$ and QF  
coefficient (\ref{eq8}), and  (d) $K_3=10^{-39}$ m$^6$/s and no QF  
LHY interaction. 
The corresponding rms sizes   $\langle x \rangle$ and $\langle z \rangle$ and the energy $E$ of these self-bound droplets are presented in Table I.
The increase of  three-body repulsion  for a fixed $a$ and $N$, from Fig.  \ref{fig5}(a)    to Fig. \ref{fig5}(b), due to an increase in $K_3$ from 0 to $10^{-39}$ m$^6$/s,
results in   larger rms sizes of the self-bound dipolar droplet, viz. Table I.   
Similarly, the use of less repulsive QF coefficient (\ref{eq8}) in Fig. \ref{fig5}(c), compared to the use of coefficient (\ref{eq7}) in   Fig. \ref{fig5}(a), has led to more attraction in the former with smaller sizes and  smaller energy. The density profiles of Figs. \ref{fig5}(a) and (d) are quite similar, which suggests that it will be reasonable to replace the QF LHY interaction by a small three-body interaction for $a\lessapprox 30a_0$ (both self-repulsive and self-attractive cases), as we will do in this paper.

In Fig. \ref{fig6} we present  1D densities $\rho_{1D}(x)$ and $\rho_{1D}(z)$ for larger values of scattering lengths ($a=60a_0,90a_0$) as well as negative values of scattering lengths. 
Although, for  $a=30a_0$, the density profiles show a sensitivity to the inclusion of a three-body interaction with an increase of length along $z$ direction, viz. Figs. \ref{fig5}(a)-(b),  this  sensitivity is much reduced for larger $a$, as illustrated in 
Figs. \ref{fig6}(d)-(e) for $a=60a_0$ and $N=10000$, where the $z$-length is practically  
unchanged after the inclusion of the three-body interaction.  For $a=90a_0$, the sensititivity to the inclusion of a three-body interaction is further reduced (not illustrated in this paper) and in Figs. \ref{fig6}(a)-(c) we display the linear densities 
for $a=90a_0, K_3=0$ and QF  LHY interaction (\ref{eq7}) for  $N=10000$, 40000 and 70000, respectively. 
The $z$-length gradually increases as $N$ is increased from 10000 to 40000 to  70000 in Figs.  \ref{fig6}(a)-(c) due to an increase of contact repulsion for 
an increased number of atoms. 
In  the self-repulsive case, the $z$-length also increases, due to an increased repulsion,  with  an increased $a$ for a fixed $N$ (=10000), viz. Figs. \ref{fig6}(a) and (d).
For smaller $a$ ($a\lessapprox 30a_0$), the imaginary part of the QF LHY interaction is large and the use of this interaction  is not recommended \cite{review}. However, it seems highly unlikely that a self-bound dipolar droplet will cease to exist in this domain and 
in that case, we employ  only the three-body interaction in the study of the formation of a dipolar droplet. In Fig. \ref{fig6}, we further display the linear densities for  $K_3= 10^{-39}$ m$^6$/s, and 
(f) $a=-a_0, N=1000$, (g) $a=-a_0, N=10000$, (h) $a=-30a_0, N=1000$, and (i)   $a=-60a_0, N=10000$, where no QF LHY interaction is included. In the self-attractive case, the $z$-length slightly decreases with  an increased $|a|$ for a fixed $N$ due to an increased attraction as can be found from Figs. \ref{fig6}(f) and (h), and 
\ref{fig6}(g) and (i); similarly, the $z$-length  increases with  an increased $N$ for a fixed $a$ ($=-a_0$) as can be found from Figs. \ref{fig6}(f) and (g).

\begin{figure}[t!]
\begin{center}
\includegraphics[width=.32\linewidth]{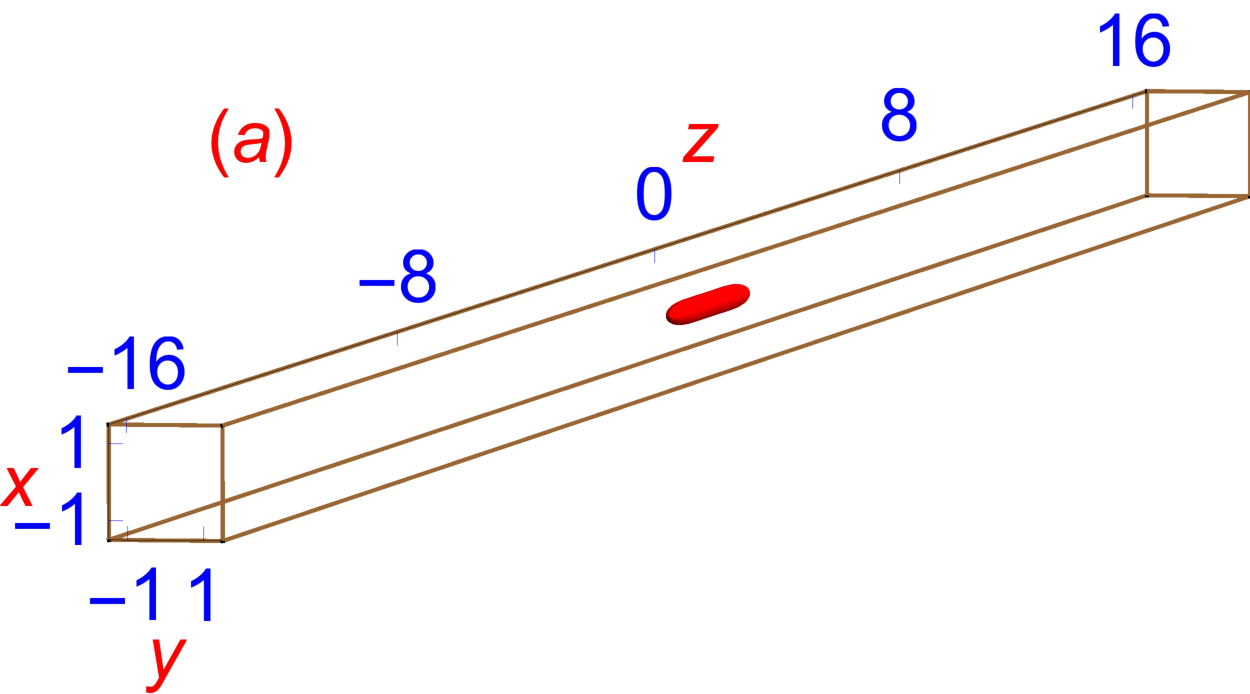}
\includegraphics[width=.32\linewidth]{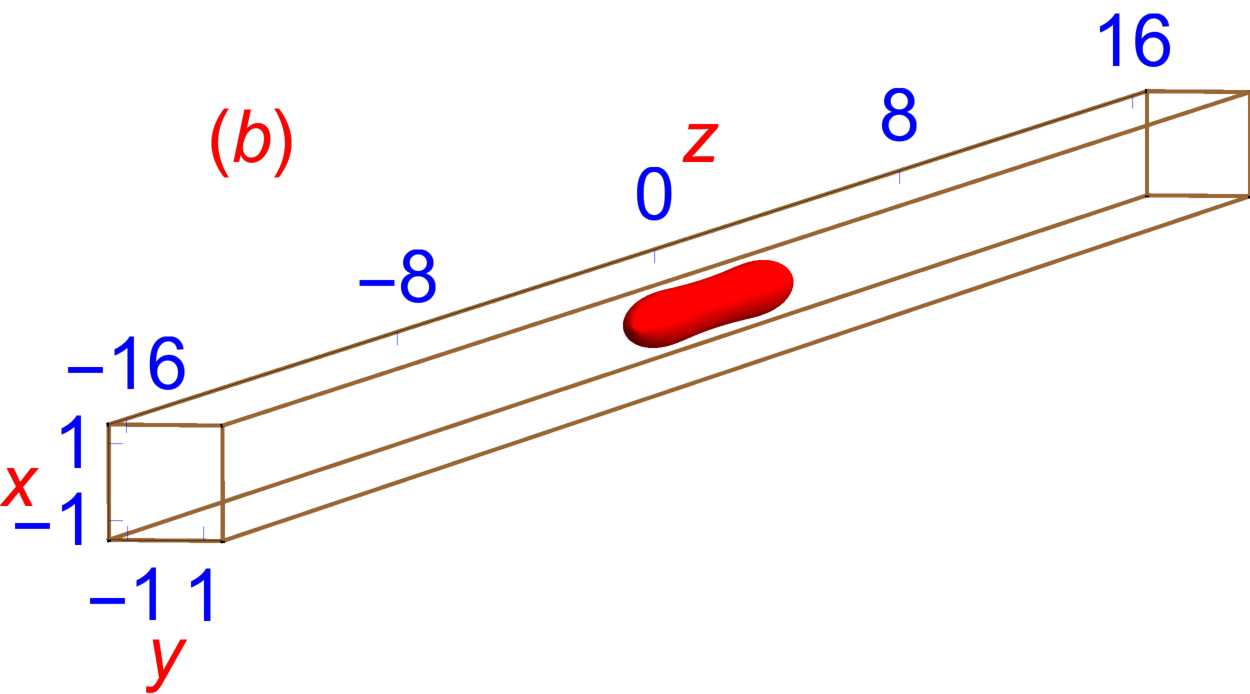}
\includegraphics[width=.32\linewidth]{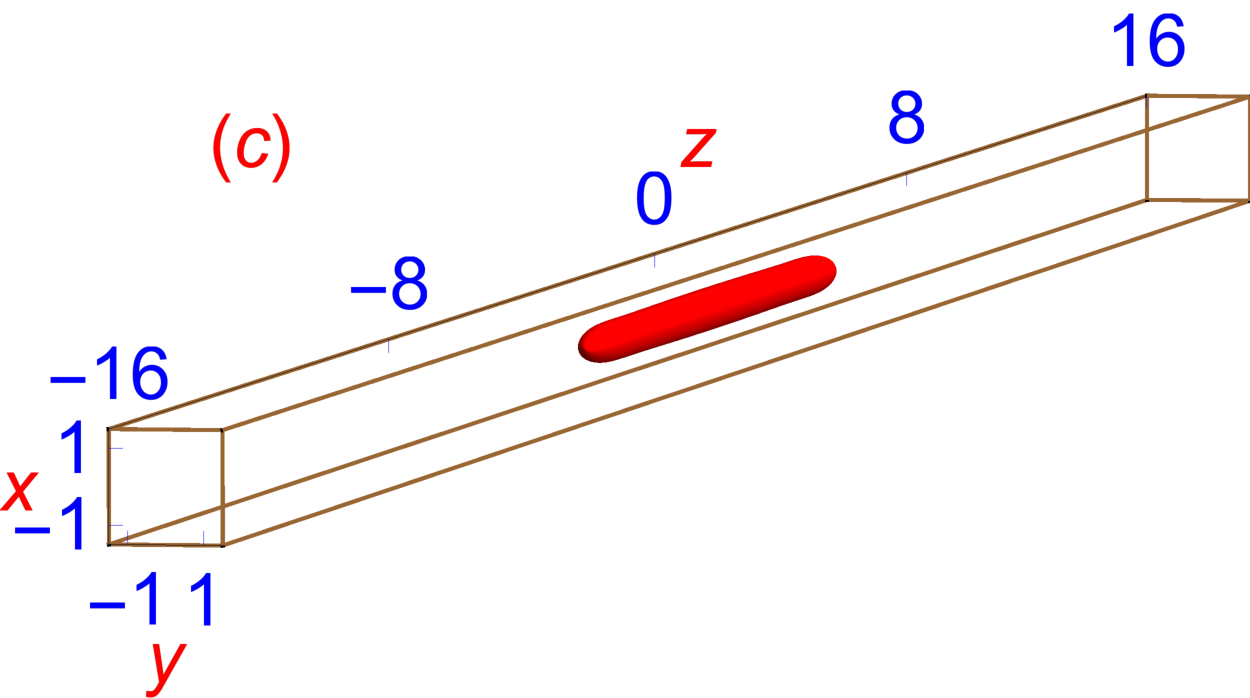}
\includegraphics[width=.32\linewidth]{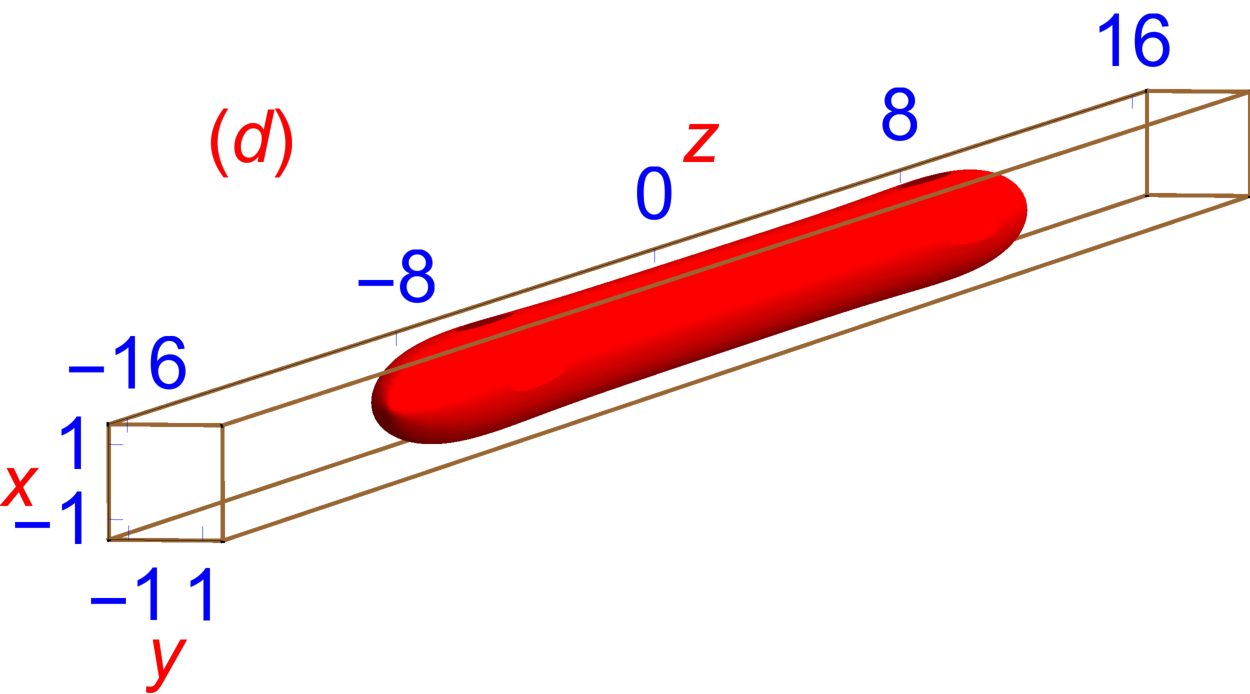}
\includegraphics[width=.32\linewidth]{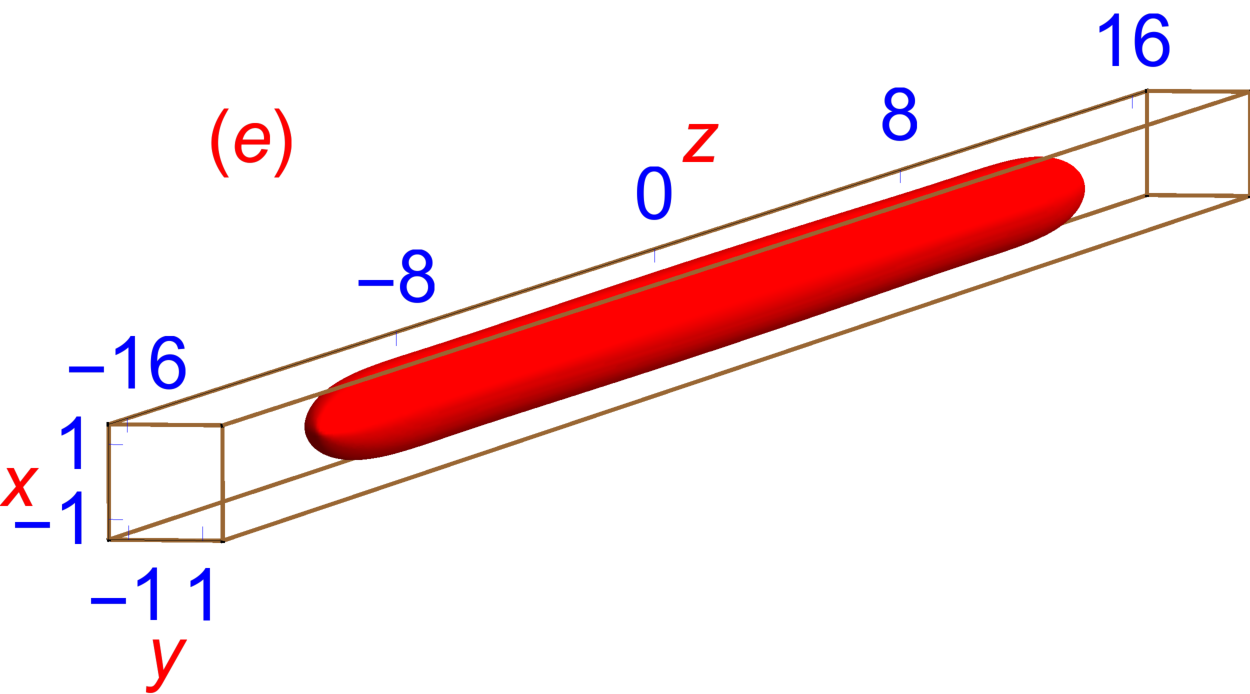}
\includegraphics[width=.32\linewidth]{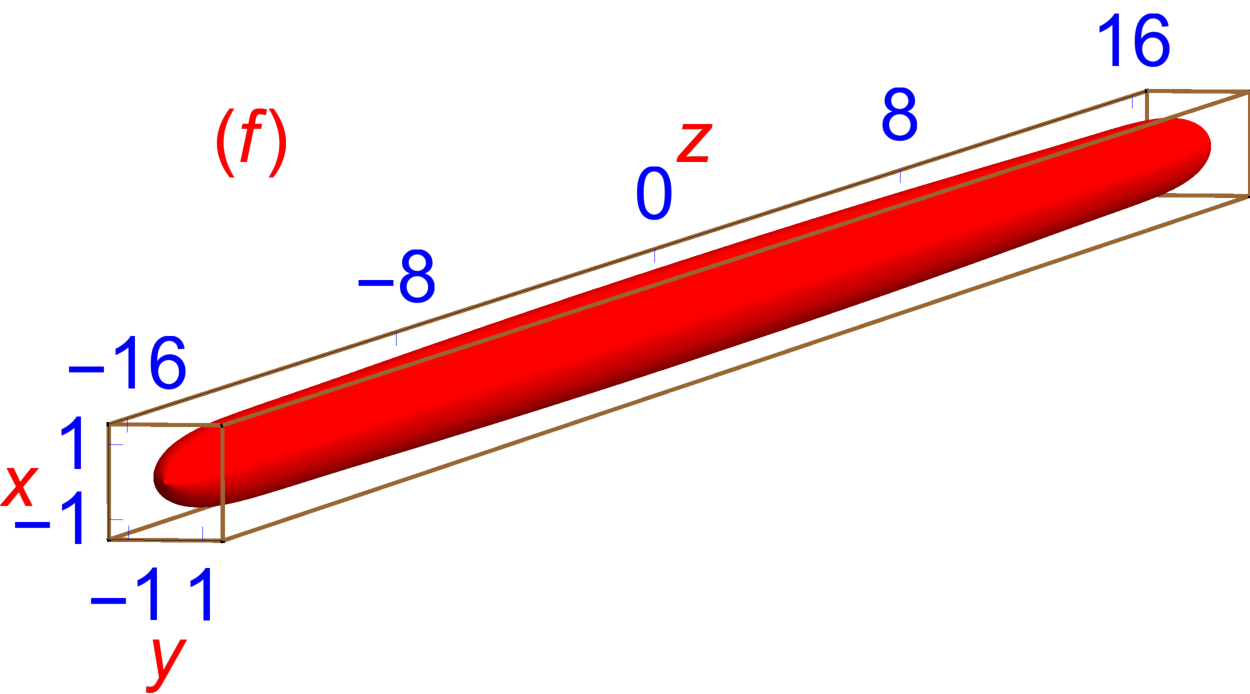}
\includegraphics[width=.32\linewidth]{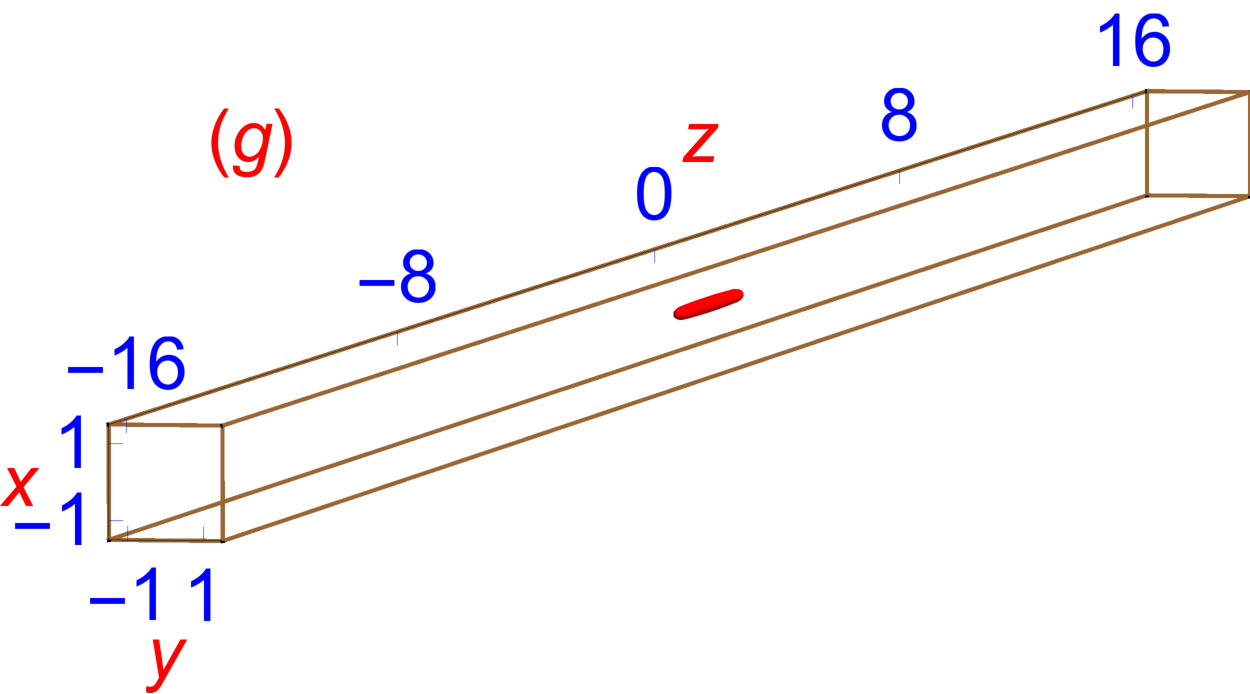}
\includegraphics[width=.32\linewidth]{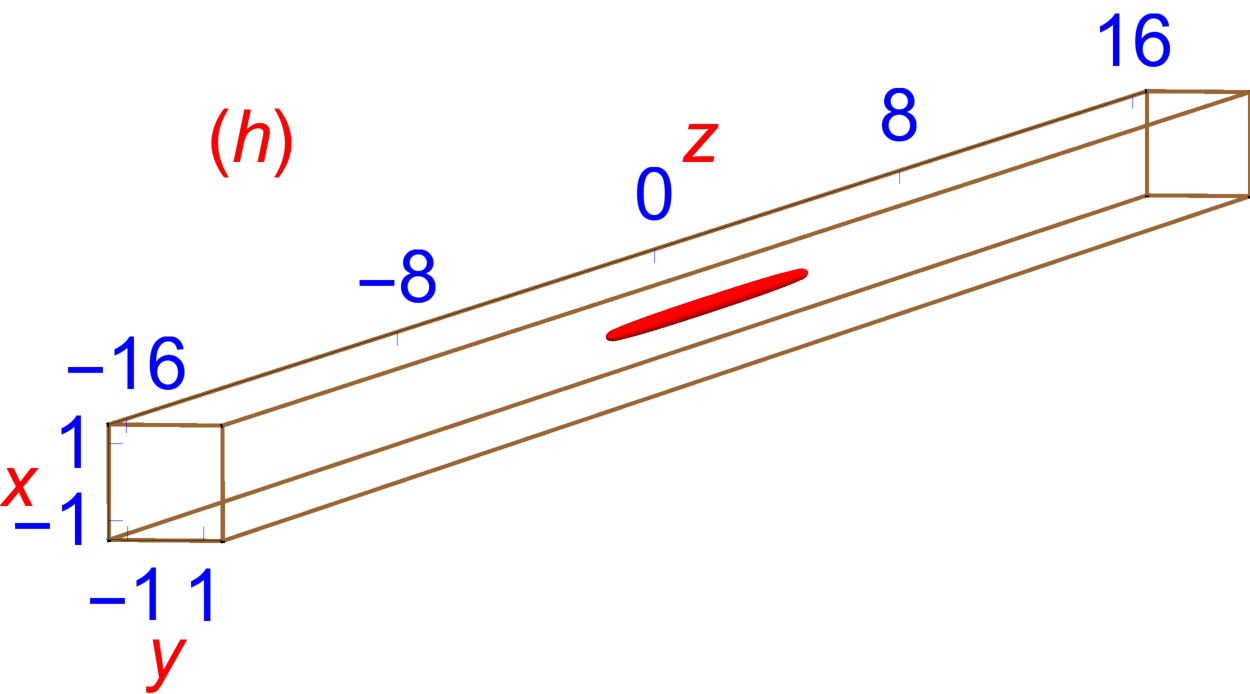}

\caption{Isodensity contour  of 3D density $|\psi(x,y,z)|^2$ for (a)  $a=30a_0, N=300$,  (b)  $a=60a_0, N=1000$,  (c)   $a=60a_0, N=10000,$ (d)   $a=90a_0, N=10000$,  (e) $a=90a_0, N=20000$,    (f) $a=90a_0, N=40000$,      (g) $a=-a_0, N=1000,$        (h) $a=-60a_0, N=10000.$     In (a)-(f) $K_3=0$ and QF coefficient (\ref{eq7}) is employed, and in (g)-(h) $K_3=10^{-39}$ m$^6$/s,
and no QF correction is used. 
The density  $|\psi(x,y,z)|^2$ on the contour is  $10^{-8}$ cm$^{-3}$ and the units of $x,y,z$ are $\mu$m.  
}

\label{fig7} 
\end{center}
\end{figure}

The change in size and shape of a self-bound dipolar droplet  with an increase of number of atoms $N$ or of scattering length $a$ is best illustrated by an   isodensity  plot of the 3D profile.   In Fig. \ref{fig7} we display the  isodensity contour of 3D density  $|\psi(x,y,z)|^2$ of a self-bound dipolar droplet for 
(a)  $a=30a_0, N=300$,    (b)  $a=60a_0, N=1000,$     (c)   $a=60a_0, N=10000$,  (d)   $a=90a_0, N=10000$,   
(e)  $a=90a_0, N=20000,$    (f)  $a=90a_0, N=40000$,    (g)   $a=-a_0, N=1000,$  (h)   $a=-60a_0, N=10000.$ 
In (a)-(f) $K_3=0$ and QF coefficient (\ref{eq7}) is employed, and in (g)-(h) $K_3=10^{-39}$ m$^6$/s,
and zero QF coefficient is used. 
The shapes are always filament-like but can be short for a small number of atoms, viz. Figs. \ref{fig7}(a)-(c),  long for a large number of atoms, viz. Figs.  \ref{fig7}(e)-(f),   fat in the self-repulsive case, viz. Figs. \ref{fig7}(d)-(f),  and becomes 
 thin as the self repulsion is reduced gradually, viz. Fig. \ref{fig7}(a), until self-attractive regime is attained,  viz. Figs. \ref{fig7}(g)-(h).
 In general, the self-repulsive  self-bound droplets are elongated along the $z$ direction and this length increases without bound with an increase of  number of atoms $N$. This can be seen from Figs. \ref{fig7}(d)-(f) for $a=90a_0$  as $N$ increases from 10000 to 20000 and then to 40000, while other parameters are held fixed. Similarly, in the self-attractive case the length also increases with $N$, with other parameters held fixed (not illustrated here).

\section{Summary}
\label{IV}
  
We studied  the formation of trapless self-bound   self-repulsive ($a>0$) and self-attractive ($a < 0$)
dipolar  droplets in 3D in  a strongly dipolar    
BEC of $^{164}$Dy atoms of dipolar length $a_{\mathrm{dd}}=130.8a_0$.
We used a beyond-mean-field model including the QF LHY 
interaction and 
a three-body interaction in addition to the usual two-body atomic contact and dipolar interactions of the mean-field GP equation. The  usual two-body  contact and dipolar interactions    
 lead to a  cubic nonlinearity in the mean-field GP  equation.  In 3D,  the attractive cubic nonlinearity in the mean-field model leads to collapse and one cannot have  a self-bound state \cite{towne}.  But due to the beyond-mean-field QF  LHY  and/or three-body interactions with  higher-order quartic  and quintic nonlinearities, respectively,  the collapse could be avoided and a strongly-dipolar BEC droplet can be formed.  
For $ 30a_0 \lessapprox a < a_{\mathrm{dd}}$, the imaginary part of the  QF  LHY interaction is small  and both the  LHY  and/or three-body interactions can be used, as appropriate. But for $a \lessapprox 30a_0$,
the imaginary part of the LHY interaction becomes large and its use is not recommended \cite{review}; in this region only the three-body interaction is used.
 
  We consider a numerical solution  and an  analytic variational approximation of the beyond-mean-field model.  The variational approximation with Gaussian ansatz for the wave function provides an analytic understanding of the formation of a strongly dipolar   self-bound droplet. Such a droplet can be formed for number of atoms $N$ larger than a critical value $N_{\mathrm{cr}}$: $N>N_{\mathrm{cr}}$. This critical number   $N_{\mathrm{cr}}$ is a monotonically increasing function of the atomic scattering length $a$ ($K_3$)  for a fixed three-body coefficient $K_3$  
($a$).  The increase of $a$ and $K_3$ enhances the repulsive interaction in the system resulting in larger   sizes of the self-bound droplet and also an increase in its negative energy.  For a  fixed $N$ and $K_3$, the negative energy of the system increases with $a$; for $N=N_{\mathrm{cr}}$ this energy becomes zero.  For   $N< N_{\mathrm{cr}}$, the energy is positive and it is not possible to form a self-bound droplet.
We presented numerical  results for the integrated reduced densities (\ref{1dx}) and (\ref{1dz}) as well as the 3D isodensity contours of the self-bound dipolar droplets in good agreement with the analytic variational results.
The 3D isodensity contours reveal that the shape of the droplet is always filament-like but can be long for a large number of atoms,  short for a small number of atoms, fat for self-repulsive droplets  or thin for self-attractive droplets. 
In view of the recent observation of trapped dipolar droplets in BECs of $^{164}$Dy \cite{1d1} and $^{168}$Er \cite{20} atoms, and of free-space binary self-bound droplets \cite{obser}, the experimental observation of the present untrapped dipolar droplets in free space seems possible, with present knowhow, after a slow removal of the traps of a trapped dipolar droplet as was emphasized  in Ref. \cite{blakie1}. 

\vskip 1 cm

\section*{Credit author statement}

Both authors were responsible for Conceptualization, Methodology, and Software. L. E. Young-S. was responsible for Formal Analysis, Investigation, Data Curation and Writing - Original Draft. S. K. Adhikari was responsible for Writing - Review and Editing and Supervision.

 \section*{Acknowledgments}
 
S.K.A. acknowledges support by the CNPq (Brazil) grant 301324/2019-0, and by the ICTP-SAIFR-FAPESP (Brazil) grant 2016/01343-7.



\begin{thebibliography}{99}






\bibitem{cite1}Y. S. Kivshar, B. A.  Malomed,  Rev. Mod. Phys. { 61}, 763  (1989);

V. S. Bagnato,  D. J. Frantzeskakis,  P. G. Kevrekidis, B. A. Malomed,  D. Mihalache,  Rom.
Rep. Phys. { 67}, 5   (2015); D. Mihalache, Rom. J. Phys. { 59}, 295  (2014).


\bibitem{cite2} Y. S. Kivshar,   G. Agrawal,   Optical Solitons: From Fibers to Photonic Crystals, (Academic
Press, San Diego, 2003).


\bibitem{cite3}K. E. Strecker, G. B. Partridge,  A. G. Truscott,   R. G. Hulet, Nature (London) { 417},  150 (2002); 

L. Khaykovich, F. Schreck, G. Ferrari, T. Bourdel, J. Cubizolles,  L. D.  Carr, Y. Castin,  C. Salomon,
  Science { 256}, 1290 (2002);

S. L. Cornish,  S. T.  Thompson,  C. E. Wieman,  Phys. Rev. Lett.  { 96}, 170401 (2006).


\bibitem{cite4}V. M. P\'erez-Garc\'ia, H. Michinel,  H. Herrero, Phys. Rev. A { 57}, 3837 (1998).


\bibitem{towne} R.Y. Chiao, E. Garmire,  C.H. Townes,  Phys. Rev. Lett. { 13}, 479
(1964).




\bibitem{SO-sol}S. K. Adhikari, Phys. Rev. A { 103}, L011301 (2021); 

H. Sakaguchi, B. Li,  B. A. Malomed, Phys. Rev. E
{ 89}, 032920 (2014);  

 S. Gautam,  S. K. Adhikari, Phys. Rev. A { 95}, 013608
(2017);

 Y.-C. Zhang, Z.-W. Zhou, B. A. Malomed,  H. Pu,
Phys. Rev. Lett. { 115}, 253902 (2015); 

 S. Gautam, S. K. Adhikari, Phys. Rev. A { 97}, 013629
(2018).

\bibitem{saka}H. Sakaguchi,  B. A. Malomed, Phys. Rev. E
{ 90}, 062922 (2014).

\bibitem{lhy}T. D. Lee, K. Huang,  C. N. Yang, Phys. Rev. { 106}, 1135
(1957).
 
\bibitem{3bdth} H.-W. Hammer, A. Nogga,  A. Schwenk, Rev. Mod. Phys. { 85}, 197 (2013).


\bibitem{3bd}A. Bulgac, Phys. Rev. Lett.  { 89}, 050402 (2002); 

 D. S. Petrov, Phys. Rev. Lett. { 112}, 103201 (2014).



\bibitem{petrov}D. S.  Petrov,  Phys. Rev. Lett. { 115}, 155302 (2015).





\bibitem{ad-bf}S. K. Adhikari, Laser Phys. Lett { 15}, 095501 (2018).



\bibitem{obser}C. R. Cabrera, L. Tanzi, J. Sanz,  B. Naylor, P. Thomas, P. Cheiney,   L. Tarruell,  Science
{ 359},  301 (2018); 


G.  Semeghini, G. Ferioli, L. Masi, C. Mazzinghi,  L. Wolswijk, F. Minardi,  M. Modugno, G. Modugno,
 M.  Inguscio,   M. Fattori,  Phys. Rev. Lett. { 120},  235301 (2018).








\bibitem{3bdex}
S. Will, T. Best, U. Schneider, L. Hackerm\"uller, D. S. L\"uhmann,  I. Bloch, Nature (London) { 465}, 197 (2010);


A. J. Daley,  J. Simon, Phys. Rev. A { 89}, 053619 (2014).

\bibitem{3bdtomio}F. Kh. Abdullaev, A. Gammal, Lauro Tomio,  T. Frederico,
Phys. Rev. A { 63}, 043604 (2001); 

H. Al-Jibbouri, I. Vidanovi\'c, Antun Bala\v z,  A. Pelster,
J. Phys. B { 46}, 065303 (2013); 

H-C Li, K-J. Chen,  J-K Xue, Chin. Phys. Lett. { 27}, 030304 (2010);


 P. Ping,  L. Guan-Qiang, Chin. Phys. B { 18}, 3221 (2009); 

M. S. Mashayekhi, J.-S. Bernier, D. Borzov, J.-L. Song,  F. Zhou, Phys. Rev. Lett. { 110}, 145301 (2013); 

S. Mostafa Moniri, H. Yavari,  E. Darsheshdar; arXiv:2109.08366. 
     

\bibitem{3bdol} A. J. Daley, J. M. Taylor, S. Diehl, M. Baranov,  P. Zoller, Phys. Rev. Lett. { 102}, 040402 (2009);

 L. Mazza, M. Rizzi, M. Lewenstein,  J. I. Cirac, Phys. Rev. A { 82}, 043629 (2010);

 M. Singh, A. Dhar, T. Mishra, R. V. Pai, B. P. Das, Phys. Rev. A { 85}, 051604(R) (2012).




\bibitem{19} M. Schmitt, M. Wenzel, F. B\"ottcher, I. Ferrier-Barbut,
 T. Pfau, Nature (London) { 539}, 259 (2016).


\bibitem{1d1}I. Ferrier-Barbut, H. Kadau, M. Schmitt, M. Wenzel,  T. Pfau, Phys. Rev. Lett. { 116}, 215301 (2016).

 
\bibitem{20}L. Chomaz, S. Baier, D. Petter, M. J. Mark, F. W\"achtler, L. Santos,  F. Ferlaino,
Phys. Rev. X { 6}, 041039 (2016).




\bibitem{qf1}A. R. P.  Lima,   A. Pelster,   Phys. Rev. A { 84}, 041604(R) (2011);

A. R. P.  Lima,   A. Pelster,   Phys. Rev. A { 86}, 063609 (2012).

\bibitem{santos}F. W\"achtler,   L. Santos, Phys. Rev. A { 93}, 061603(R) (2016).
 

\bibitem{blakie1}D. Baillie, R. M. Wilson, R. N. Bisset,  P. B. Blakie, Phys. Rev. A { 94}, 021602(R) (2016).
\bibitem{blakie2} 
D. Baillie, R. M. Wilson,  P. B. Blakie, Phys. Rev. Lett. { 119}, 255302 (2017);

S. Pal, D. Baillie,  P. B. Blakie, Phys. Rev. A { 105}, 023308 (2022).

\bibitem{drop3}F. W\"achtler,  L.  Santos,   Phys. Rev. A { 94}, 043618 (2016).


 \bibitem{adhidrop}
Z.-K. Lu, Y. Li, D. S. Petrov,  G. V. Shlyapnikov, Phys. Rev. Lett. { 115}, 075303 (2015).


\bibitem{sabri}T. Ramakrishnan,   S. Subramaniyan, Phys. Lett. A { 383}, 2033 (2019).



\bibitem{review}L. Chomaz, I. Ferrier-Barbut, F. Ferlaino, B. Laburthe-Tolra, B. L. Lev, 
 T. Pfau,  arXiv:2201.02672.


\bibitem{dipbec} T Lahaye, C Menotti, L Santos, M Lewenstein,  T Pfau,
Rep. Prog. Phys. { 72}, 126401 (2009).






\bibitem{dip} R. Kishor Kumar, L. E. Young-S., D. Vudragovi\'c, A. Bala\v{z}, P. Muruganandam,  S. K. Adhikari, Comput. Phys. Commun. { 195}, 117 (2015).


\bibitem{yuka}V. I. Yukalov, Laser Phys.  { 28}, 053001 (2018).


\bibitem{expdy}Y. Tang, A. Sykes, N. Q. Burdick, J. L. Bohn,  B. L. Lev,
Phys. Rev. A { 92}, 022703 (2015).

 

\bibitem{muntsa}M. Abad, M. Guilleumas, R. Mayol, M. Pi,  D. M. Jezek, Phys. Rev. A { 79}, 063622 
(2009);

 M. Abad, M. Guilleumas, R. Mayol, M. Pi,  D. M. Jezek,
Phys. Rev. A { 81}, 043619 (2010).






 \bibitem{blakie}R. N. Bisset, R. M. Wilson, D. Baillie,  P. B. Blakie,
Phys. Rev. A { 94}, 033619 (2016).


\bibitem{3bdyexp}J. S\"oding, D. Gu\'ery-Odelin, P. Desbiolles, F. Chevy, H. Inamori,   J. Dalibard, 
Applied Phys. B { 69}, 257 (1999);

D. M. Stamper-Kurn, M. R. Andrews, A. P. Chikkatur, S. Inouye, H.-J. Miesner, J. Stenger,  W. Ketterle,
Phys. Rev. Lett. { 80}, 2027 (1998).



\bibitem{pohl}Y.-C. Zhang, F. Maucher,  T. Pohl,
Phys. Rev. Lett. { 123}, 015301 (2019);

Y.-C. Zhang, T. Pohl, F. Maucher, Phys. Rev. A { 104},
013310 (2021).


\bibitem{pfau}J. Hertkorn, J.-N. Schmidt, M. Guo, F. B\"ottcher, K.S.H. Ng, S.D. Graham, P. Uerlings, T. Langen, M. Zwierlein,  T. Pfau, Phys. Rev. Research { 3}, 033125 (2021); 


F. B\"ottcher, J.-N. Schmidt, J. Hertkorn, K. S. H. Ng, S. D. Graham, M. Guo, T. Langen,  T. Pfau,
 Rep. Prog. Phys. { 84} 012403 (2021).


\bibitem{crank}P. Muruganandam,  S. K. Adhikari, Comput. Phys. Commun. { 180}, 1888       (2009).

\bibitem{omp}V. Lon\v car, L. E. Young-S., S. \v Skrbi\'c, P. Muruganandam, S. K. Adhikari, A. Bala\v z,
Comput. Phys. Commun. { 209},   190 (2016).

\bibitem{feshbach} S. Inouye, M.R. Andrews, J. Stenger, H.J. Miesner, D.M.
Stamper-Kurn,  W. Ketterle, Nature (London) { 392},
151 (1998);

 C. Chin, R.Grimm,  P. Julienne,   E. Tiesinga,   Rev. Mod. Phys. { 82}, 1225 (2010).

\bibitem{32} F. K. Fatemi, K. M.  Jones,   P. D.   Lett,  Phys. Rev. Lett. { 85}, 4462
(2000).




\bibitem{K3vary} A. Hammond, L. Lavoine,  T. Bourdel,  	arXiv:2112.01782;

   F. M. Gambetta,  C. Zhang, M. Hennrich,
I. Lesanovsky,  W. Li, Phys. Rev. Lett. { 125}, 133602 (2020).



 


 



 





 



 
 
\end{thebibliography}
\end{document}